\documentclass[journal, final, twoside]{IEEEtran}
\usepackage{amsthm, amsmath, amssymb, bm}
\usepackage{enumerate, graphicx, cite, color, soul}
\usepackage{algorithm}
\usepackage[noend]{algpseudocode}

\newcommand{\ee}{\ensuremath{{\rm e}}}

\renewcommand{\leq}{\ensuremath{\leqslant}}
\renewcommand{\geq}{\ensuremath{\geqslant}}
\newcommand{\inv}[1]{\ensuremath{\frac{1}{#1}}}
\newcommand{\abs}[1]{\ensuremath{\left| #1 \right|}}

\newcommand{\CCal}{\ensuremath{\set{C}}}
\newcommand{\DCal}{\ensuremath{\set{D}}}

\newcommand{\HCal}{\ensuremath{\set{H}}}

\newcommand{\SCal}{\ensuremath{\set{S}}}
\newcommand{\VCal}{\ensuremath{\set{V}}}
\newcommand{\Rbb}{\ensuremath{\mathbb{R}}} 
 
\newcommand{\Nbb}{\ensuremath{\mathbb{N}}} 
\newcommand{\Sbb}{\ensuremath{\mathbb{S}}} 
\newcommand{\set}[1]{\ensuremath{\mathcal{#1}}}
\renewcommand{\vec}[1]{\ensuremath{\bm{#1}}}
\newcommand{\adjoint}{\ensuremath{{\intercal}}}
\newcommand{\ma}[1]{\ensuremath{\mathsf{#1}}}
\newcommand{\norm}[1]{\ensuremath{\left\| #1\right\|}}
\newcommand{\scp}[2]{\ensuremath{\left\langle #1, #2 \right\rangle}}
\newcommand{\Ebb}{\ensuremath{\mathbb{E}}} 
\newcommand{\Prob}{\ensuremath{\mathbb{P}}} 
\DeclareMathOperator*{\argmin}{argmin}

\newcommand{\ie}{\textit{i.e.}}
\newcommand{\eg}{\textit{e.g.}}
\newcommand{\etal}{\textit{et al.}}
\newcommand{\refeq}[1]{(\ref{#1})}
\newcommand{\add}[1]{{\color{black} #1}}
\newtheorem{theorem}{Theorem}[section]
\newtheorem{lemma}[theorem]{Lemma}

\newtheorem{definition}[theorem]{Definition}
\newtheorem{assumption}{Assumption}

\newcommand{\longtitle}{{Recipes for stable linear embeddings from Hilbert spaces to $\Rbb^m$}}
\newcommand{\shorttitle}{{Recipes for stable linear embeddings from Hilbert spaces to $\Rbb^m$}}
\newcommand{\GPlong}{{Gilles~Puy}} \newcommand{\GPshort}{{G.~Puy}}
\newcommand{\RGlong}{{R\'emi~Gribonval}} \newcommand{\RGshort}{{R.~Gribonval}}
\newcommand{\MDlong}{{Mike~E.~Davies}} \newcommand{\MDshort}{{M.~E.~Davies}}
\newcommand{\ERC}{{This work was partly funded by the European Research Council, PLEASE project (ERC-StG-2011-277906). \MDshort\ would like to acknowledge the support of EPSRC grant EP/J015180/1}}
\newcommand{\refMainTheorem}{\ref{th:main_theorem}}
\makeatletter
\renewcommand*{\ALG@name}{Recipe}
\makeatother

\title{\longtitle}
\author{
\GPlong, \MDlong, and \RGlong 
\thanks{\GPshort\ is with {Technicolor}, {975 Avenue des Champs Blancs, 35576 Cesson-S\'evign\'e, FR. He was with INRIA Rennes - Bretagne Atlantique, Campus de Beaulieu, 35042 Rennes Cedex, FR, when the first verison of this work was submitted.}}
\thanks{\MDshort\ is with the institute for Digital Communications (IDCom), The University of Edinburgh, EH9 3JL, UK.}
\thanks{\RGshort\ is with  with INRIA Rennes - Bretagne Atlantique, Campus de Beaulieu, 35042 Rennes Cedex, FR.}
\thanks{\ERC.}
\thanks{This work is an extension of the paper \cite{puy15} presented at the European Signal Processing Conference (EUSIPCO), 2015.}
}

\begin{document}

\maketitle

\begin{abstract}
\add{We consider the problem of constructing a linear map from a Hilbert space $\HCal$ (possibly infinite dimensional) to $\Rbb^m$ that satisfies a restricted isometry property (RIP) on an arbitrary signal model, i.e., a subset of $\HCal$. We present a generic framework that handles a large class of low-dimensional subsets but also \emph{unstructured} and \emph{structured} linear maps. We provide a simple recipe to prove that a random linear map satisfies a general RIP with high probability. We also describe a generic technique to construct linear maps that satisfy the RIP. Finally, we detail how to use our results in several examples, which allow us to recover and extend many known compressive sampling results.}
\end{abstract}

\IEEEpeerreviewmaketitle

\section{Introduction}

\IEEEPARstart{T}{he} restricted isometry property (RIP) is at the core of many theoretical developments in compressive sensing (CS). It allows ones to show that sparse signals can be ``captured'' by few linear and non-adaptive measurements and recovered by non-linear decoders \cite{foucart13}. In a finite dimensional space, a matrix $\ma{A} \in \Rbb^{m \times n}$ satisfies the RIP on a general set $\mathcal{S} \subset \Rbb^n$, if there exists a constant $\delta \in (0, 1)$, such that for all $x \in \mathcal{S}$,
\begin{align}
\label{eq:RIP1}
(1 - \delta) \norm{x}_2^2 \leq \norm{\ma{A} x}_2^2 \leq (1 + \delta) \norm{x}_2^2.
\end{align}
Random matrices with independent entries drawn from the centered Gaussian distribution with variance $m^{-1}$ are examples of matrices that satisfy the RIP with high probability for many low-dimensional sets such as \add{$\mathcal{S} = \Sigma_{2k}$, associated to $k$-sparse signals \cite{foucart13}, $\mathcal{S} = \Sigma_{2r}$, associated to rank-$r$ matrices \cite{candes11a}, or $\mathcal{S} = \{x_1-x_2 \vert x_1, x_2 \in \Sigma\}$, associated to a compact Riemannian manifold $\Sigma$ \cite{eftekhari14}. In these scenarios, the RIP holds for a number of measurements $m$ essentially proportional to a measure of intrinsic dimension of $\mathcal{S}$.}

\add{In this work, we extend the construction of linear maps that satisfy the RIP in a finite-dimensional ambient space to linear maps that satisfy the RIP on subsets of a possibly \emph{infinite}-dimensional space. This extension of the CS theory to infinite-dimensional spaces is important to properly apply CS in an analog setting \cite{adcock14}, explore connections with the sampling of signals with finite rate of innovation \cite{vetterli02}, or also in machine learning to develop efficient methods to compute information-preserving sketches of probability distributions \cite{thaper02, bourrier13}. As an example of the application of our results, we will explain how to build a stable linear embedding of sparse signals in the Haar wavelet basis of $L_2([0, 1])$ with a sampling in the Fourier basis in Section~\ref{sec:mri_example}. Note that this type of signal model and sampling is often used to study the theoretical aspects of CS-MRI acquisition.}

\subsection{The normalised secant set}

\add{The RIP is a very convenient tool when one wants to prove that a signal $x \in \mathcal{H}$ can be reconstructed from its compressed measurements $\ma{A} x$ when it belongs to a given model set, \ie, $x \in \Sigma \subset \mathcal{H}$}. If $\ma{A}$ satisfies the RIP on the set of $2k$-sparse vectors then every $k$-sparse vector $x$ can be accurately and stably recovered from its noisy measurements $\ma{A} x + \vec{n}$ by solving the Basis Pursuit problem \cite{foucart13}. For more general low-dimensional signal models $\Sigma$, one can prove that a signal $x \in \Sigma$ can be stably recovered from its compressed measurements if the matrix $\ma{A}$ satisfies the RIP on the secant set \add{$\mathcal{S} = \Sigma - \Sigma$}, \cite{blumensath11, bourrier14}, \ie, if there exists a constant $\delta \in (0, 1)$ such that
\begin{align*}
(1 - \delta) \norm{x_1 - x_2}_2^2 \leq \norm{\ma{A} (x_1 - x_2)}_2^2 \leq (1 + \delta) \norm{x_1 - x_2}_2^2.
\end{align*}
for all $x_1, x_2 \in \Sigma$. When this condition is satisfied, we say that the matrix $\ma{A}$ stably embeds the set $\Sigma$ in $\Rbb^m$. 

Let us recall that the \emph{secant set} of $\Sigma$ is defined as
\begin{align*}
\Sigma - \Sigma := \left\{ y = x_1 - x_2 \; \Big\vert \; x_1, x_2 \in \Sigma \right\}.
\end{align*}
One can remark that the above condition for stable recovery is equivalent to
\begin{align}
\label{eq:RIP_centered}
\sup_{z \in \add{\mathcal{S}({\Sigma})}} \abs{\norm{\ma{A} z}_2^2 - 1} \leq \delta,
\end{align}
where 
\begin{align*}
\add{\mathcal{S}({\Sigma})} := \left\{ z = {y}/{\norm{y}_{2}} \; \Big\vert \; y \in (\Sigma-\Sigma) \setminus \{\vec{0}\} \right\}
\end{align*}
is the \emph{normalised secant set} of $\Sigma$. This indicates that the set of interest to prove that a matrix $\ma{A}$ stably embeds the set $\Sigma$ is not directly $\Sigma$, but rather its normalised secant set $\add{\mathcal{S}({\Sigma})}$. 

From now on, we concentrate on proving a generalised version of \refeq{eq:RIP_centered} for an arbitrary set $\SCal$ lying on the unit sphere in the ambient space. If one wants to prove that a matrix or, more generally, a linear map stably embeds a set $\Sigma$, one just needs to substitute $\mathcal{S}({\Sigma})$ for $\SCal$ in the following results. 

\add{In this paper, we consider that the ambient space is an infinite-dimensional Hilbert space denoted by $\HCal$. The inner product in $\HCal$ is denoted by $\scp{\cdot}{\cdot}$ and the associated norm by $\norm{\cdot}$. Vectors in $\HCal$ are given a bold letter while vectors in $\Rbb^n$ are given a regular letter. We denote by $\Sbb$ the unit sphere in $\HCal$. We assume everywhere $\SCal \subset \Sbb$: $\norm{\vec{x}} = 1$ for all $\vec{x} \in \SCal$.}

\subsection{Measuring the dimension of $\SCal$}

All the developments in this work are based on the assumption that $\SCal$ has a small intrinsic dimension. In the literature, several definitions of dimension exist. The reader can refer to, \eg, the monograph of Robinson \cite{robinson11}, for an exhaustive list of definitions. In an infinite-dimensional space, one should be careful with the definition of dimension used. Indeed, as described in \cite{robinson11}, there are examples of sets for which no stable linear embedding to a finite dimensional space exists even though their dimension is finite (according to some definition). Therefore, there is no hope to construct a linear map that satisfies the RIP for these sets. 

In this paper, we use the \emph{upper box-counting dimension} as our measure of dimension. The upper box-counting dimension is also at the centre of most of the developments in \cite{robinson11}.

\begin{definition}[Covering number]
Let $\epsilon > 0$. The covering number $N_{\SCal}(\epsilon)$ of $\SCal$ is the minimum number of closed balls of radius $\epsilon$, with respect to the norm $\norm{\cdot}$, with centres in $\SCal$ needed to cover $\SCal$. The set of centres of these balls is a minimal $\epsilon$-net for $\SCal$.
\end{definition}
\begin{definition}[Upper box-counting dimension]
The upper box-counting dimension of $\SCal$ is
\begin{align*}
{\rm dim}(\SCal) :=  \limsup_{\epsilon \rightarrow 0} \; {\log[N_{\SCal}(\epsilon)]}/{\log[1/\epsilon]}.
\end{align*}
\end{definition}

From the above definition, one can remark that if $d > {\rm dim}(\SCal)$ then there exists $\epsilon_{\SCal} > 0$ such that $N_{\SCal}(\epsilon) \leq \epsilon^{-d}$ for all $\epsilon \leq \epsilon_{\SCal}$. In this paper, we make the following assumption on~$\SCal$.
\begin{assumption}
\label{as:dimension}
The set $\SCal \subset \Sbb$ has a finite upper box-counting dimension ${\rm dim}(\SCal)$ which is \underline{strictly} bounded by $s\geq1$:  ${\rm dim}(\SCal)~<~s$.

Therefore, there exists a model-set dependent constant $\epsilon_{\SCal}~\in~(0, 1/2)$ such that $N_{\SCal}(\epsilon) \leq \epsilon^{-s}$ for all $\epsilon \leq \epsilon_{\SCal}$.
\end{assumption}

In many well-behaved cases, the upper-box counting dimension of the normalised secant set $\SCal = \add{\mathcal{S}({\Sigma})}$ of $\Sigma$ is simply twice the upper-box counting dimension of $\Sigma$ (or $\Sigma \cap \Sbb$ if $\Sigma$ is not bounded). For example, the set of normalised $k$-sparse vectors in $\Rbb^n$ has dimension $k$ and its secant set has dimension $2k$. The same results holds for low-rank matrices and low-dimensional manifolds. However, this is not always the case as we will see in Section \ref{sec:counterexample}.

\subsection{Contributions and organisation of the paper}

This work has four main contributions.

\begin{itemize}
\item First, given a set $\SCal$ that satisfies Assumption \ref{as:dimension} and a random linear map $L: \HCal \rightarrow \Rbb^m$, we factorise the proof of the RIP for $L$ on $\SCal$ into a small number of steps. We separate the steps that are problem-specific from the ones that are common to most problems and do not need to be repeated every time. This allows us to provide a simple recipe for the proof of the RIP, with techniques similar to those of, \eg, \cite{talagrand96, clarkson08, rauhut10, dirksen14b, eftekhari14}. This is the subject of Section \ref{sec:recipe}.
\item Second, given an \emph{arbitrary} set $\SCal$ that satisfies Assumption \ref{as:dimension}, we propose a \emph{generic construction} of linear maps that satisfy the RIP on $\SCal$. This construction is made of two steps. In the first step, we show that one can always find a finite-dimensional subspace of large but finite dimension that accurately approximates $\SCal$. We then use this subspace to build a linear functional that maps the vectors in $\SCal$ into finite but potentially large dimension while preserving their norm to a prescribed accuracy. The second step consists in reducing the embedding dimension further with a random matrix. This construction is detailed in Section \ref{sec:generic_linear_map}.
\item Third, we show how to use the developed techniques to recover and extend many known CS results. In particular, we show that our recipe is general enough to handle \emph{structured} measurement strategies such as the ones proposed in \cite{cai15, chen15}. These examples are presented in Section \ref{sec:examples}.
\item Fourth, we show that while it is sufficient to have a set $\SCal$ of finite upper-box counting dimension to guarantee the existence of linear maps that have the RIP on $\SCal$, this condition is not \emph{necessary}. This fact is discussed in Section \ref{sec:counterexample}.
\end{itemize}

We discuss related works in Section \ref{sec:related_works} and conclude in Section \ref{sec:conclusion}. All technical proofs can be found in the appendices.

%
\section{Recipe to prove that the RIP holds}
\label{sec:recipe}

In this section, we give a generic recipe to prove that a random linear map $L : \HCal \rightarrow \Rbb^m$ preserves the norm of all vectors in the set $\SCal \subset \Sbb$. 

We suppose that $L$ is built by drawing at random $m$ vectors $(\vec{l}_1, \ldots, \vec{l}_m)$ in $\HCal^m$ using a probability measure $\mu$ on $\HCal^{m}$: 
\begin{align}
\label{eq:generic_linear_map}
L \colon \HCal 	& \longrightarrow  	\Rbb^m \nonumber \\
\vec{x} 	& \longmapsto 	(\scp{\vec{l}_i}{\vec{x}})_{1 \leq i \leq m}.
\end{align}
Notice that $L$ is a continuous linear map.

In the following, we start by introducing a generalised form of RIP. Then we detail generic properties of the probability measure $\mu$ which are sufficient to prove that $L$ preserves the norm of all vectors in $\SCal$ with high probability, provided that $m$ is sufficiently large. Finally, we give a recipe to show that a random linear map preserves the norm of all vectors in $\SCal$ with high probability.

\subsection{Generalised form of the RIP}

Let us come back to the usual form of the RIP~\refeq{eq:RIP1}, which involves the Euclidian norms in the ambient spaces. First, we notice that it can be rewritten in the following equivalent form:
\begin{align}
\label{eq:RIP2}
\abs{\norm{\ma{A} \vec{x}}_2^2 - \norm{\vec{x}}_2^2} \leq \delta \norm{\vec{x}}_2^2.
\end{align}
Second, we remark that for typical random matrices $\ma{A}$ satisfying this RIP, one has $\Ebb\norm{\ma{A} \vec{x}}_2^2 = \norm{\vec{x}}_2^2$. This property is satisfied, \eg, for random matrices $\ma{A} \in \Rbb^{m \times n}$  whose entries are independent Gaussian variables with  zero-mean and variance $1/m$, for $\pm 1/\sqrt{m}$ Bernoulli random matrices, or for (rescaled) sensing matrices constructed by selecting independently $m$ vectors from an orthonormal basis using the uniform distribution. For all these cases, inequality \refeq{eq:RIP2} becomes
\begin{align}
\label{eq:RIP3}
\abs{\norm{\ma{A} \vec{x}}_2^2 - \Ebb\norm{\ma{A} \vec{x}}_2^2} \leq \delta \norm{\vec{x}}_2^2.
\end{align}
In this form, we see that the right-hand side of \refeq{eq:RIP3} characterises how much $\norm{\ma{A} \vec{x}}_2^2$ deviates from its mean. In \refeq{eq:RIP3}, this deviation is proportional to $\norm{\vec{x}}_2^2$, the square of the Euclidean norm in the ambient space. We replace here the Euclidean norm by the norm $\norm{\cdot}$ in the considered Hilbert space. 
Finally, instead of concentrating only on the $\ell_2$-norm in the measurement space, we will consider arbitrary $\ell_p$-norms, $p\geq1$.

We define the following semi-norm for all $\vec{x} \in \HCal$,
\begin{align*}
\norm{\vec{x}}_{\mu, p} := \left(\Ebb_{\mu} \norm{L(\vec{x})}_p^p\right)^{1/p}.
\end{align*}
Gathering all the above remarks and adapting them to the case of random linear maps from $\HCal$ to $\Rbb^m$, we consider  the following general form for the desired RIP:
\begin{align*}
\abs{\norm{L(\vec{x})}_p^p - \norm{\vec{x}}_{\mu, p}^p} \leq \delta, \quad \forall \vec{x} \in \SCal \subset \Sbb.
\end{align*}
\begin{definition}[RIP]
Let $L~:~\HCal \rightarrow \Rbb^m$ be a linear map. Define
\begin{align}
\label{eq:true_RIP_constant}
\delta_{\SCal, \mu, p} := \sup_{\vec{x} \in \SCal} \abs{\norm{L(\vec{x})}_p^p - \norm{\vec{x}}_{\mu, p}^p}
\end{align}
and
\begin{align*}
\underline{\delta}_{\SCal, \mu, p} := \inf_{\vec{x} \in \SCal} \norm{\vec{x}}_{\mu, p}^p.
\end{align*}

The linear map $L$ satisfies the RIP on $\SCal \subset \Sbb$ with constant $\delta$ if $ \delta_{\SCal, \mu, p} \leq \delta < \underline{\delta}_{\SCal, \mu, p}$.
\end{definition}

It will also be convenient to define
\begin{align*}
\bar{\delta}_{\SCal, \mu, p} := \sup_{\vec{x} \in \SCal} \norm{\vec{x}}_{\mu, p}^p.
\end{align*}
To simplify notations, we substitute $\delta_{p}$,  $\underline{\delta}_{p}$, and $\bar{\delta}_{p}$ for $\delta_{\SCal, \mu, p}$, $\underline{\delta}_{\SCal, \mu, p} $, and $\bar{\delta}_{\SCal, \mu, p}$, respectively, but one should keep in mind that these quantities depend on $\SCal$, $p$ and $\mu$.

We remark that if $L$ satisfies the RIP on $\SCal$ with constant $\delta$ then 
\begin{align*}
\underline{\delta}_p - \delta \leq \norm{L(\vec{x})}_p^p \leq \bar{\delta}_p - \delta,
\end{align*}
for all $\vec{x} \in \SCal$. The condition $\delta < \underline{\delta}_p$ ensures that no vector in $\vec{x} \in \SCal$ is in the null space of $L$. Indeed, we have
\begin{align*}
\norm{L(\vec{x})}_p^p \geq \underline{\delta}_p - \delta > 0,
\end{align*}
for all $\vec{x} \in \SCal$. 

In the remaining part of this section, we give generic sufficient conditions to ensure that the RIP holds with high probability. We will see how to recover classical RIP results in finite dimensions with $p=2$ in Section \ref{sec:rip_hilbert}.

\subsection{Concentration inequalities and main result}

If Assumption \ref{as:dimension} holds, the only other ingredients needed to prove that a random linear map $L$ of the form of \refeq{eq:generic_linear_map} satisfies the RIP with high probability are concentration inequalities. The choice of the probability distribution $\mu$ is thus important to ensure a preservation of the norm of the vectors in $\SCal$ by $L$. In this section, we assume that the following general concentration inequalities hold.

\begin{assumption}
\label{as:concentration}
Define the function
\begin{align*}
h_{L, \mu, p} \colon \HCal 	& \longrightarrow  	\Rbb \nonumber \\
 \vec{x} 	& \longmapsto 	\norm{L(\vec{x})}_p^p - \norm{\vec{x}}_{\mu, p}^p.
\end{align*}
There exist two constants $c_1, c_2 \in (0, \infty]$ such that for any fixed $\vec{y}, \vec{z} \in \SCal \cup \{ \vec{0} \}$,
\begin{align}
\label{eq:prob_bound_increment_1}
\Prob \left\{ \abs{h_{L, \mu, p}(\vec{y}) - h_{L, \mu, p}(\vec{z})} \geq \lambda \norm{\vec{y} - \vec{z}} \right\}
\leq 2 \ee^{- c_1 m \lambda^2},
\end{align}
for every $0 \leq \lambda \leq {c_2}/{c_1}$, and
\begin{align}
\label{eq:prob_bound_increment_2}
\Prob \left\{ \abs{h_{L, \mu, p}(\vec{y}) - h_{L, \mu, p}(\vec{z})} \geq \lambda \norm{\vec{y} - \vec{z}} \right\}
\leq 2  \ee^{- c_2 m \lambda},
\end{align}
for every $\lambda \geq {c_2}/{c_1}$.
\end{assumption}

To simplify notations, we substitute $h_{p}$ for $h_{L, \mu, p}$ in the remaining part of the paper.

We remark that we can have $c_1 = \infty$ or $c_2 = \infty$. This is to handle the case where one of the above bounds hold for all $\lambda \geq 0$. If $c_1 = \infty$ then \refeq{eq:prob_bound_increment_2} holds for all $\lambda \geq 0$. Similarly, if $c_2 = \infty$ then \refeq{eq:prob_bound_increment_1} holds for all $\lambda \geq 0$.

We can now state our main theorem from which all the following results are derived.
\begin{theorem}
\label{th:main_theorem}
Let $L~:~\HCal \rightarrow \Rbb^m$ be a random linear map constructed as in \refeq{eq:generic_linear_map} using a probability distribution $\mu$. If Assumption \ref{as:dimension} and Assumption \ref{as:concentration} hold, then for any $\xi, \delta \in (0, 1)$, we have: $\delta_{p} \leq \delta$ with probability at least $1 - \xi$ provided that
\begin{align}
\label{eq:cond_m_main_theorem}
m \geq \frac{C}{\min(c_1, c_2) \; \delta^2} \, \max\left\{s \log\left(\inv{\epsilon_{\SCal}}\right),\ \log\left(\frac{6}{\xi}\right)\right\},
\end{align}
where $C>0$ is an absolute constant.
\end{theorem}

Inequality \refeq{eq:cond_m_main_theorem} above involves the \emph{natural logarithm}. The proof of this theorem is available in Appendix \ref{app:main_theorem}. The proof is based on a chaining argument which is a powerful technique to obtain sharp bounds for the supremum of random processes, see, \eg, \cite{talagrand96, clarkson08, rauhut10, dirksen14b, eftekhari14}. This technique can be viewed as a refinement of the classical $\epsilon$-net argument such as used in, \eg, \cite{vershynin12, baraniuk08}.

\subsection{Proof recipe for the RIP}

We can now deduce the following recipe to show that a random linear map $L$ of the form of \refeq{eq:generic_linear_map} satisfies the RIP on~$\SCal$.

\begin{algorithm}
\caption{Recipe to prove that $L$ satisfies the RIP on $\SCal$}
\label{alg:recipe_rip}
\begin{algorithmic}[1]
\State Prove that the set $\SCal$ has finite upper box-counting dimension (Assumption \ref{as:dimension}). 
\State Compute $\underline{\delta}_p := \inf_{\vec{x} \in \SCal} \Ebb_{\mu} \norm{L(\vec{x})}_p^p$.
\State Prove that the concentration inequalities \refeq{eq:prob_bound_increment_1} and \refeq{eq:prob_bound_increment_2} hold (Assumption \ref{as:concentration}).
\State Choose $m$ such that \refeq{eq:cond_m_main_theorem} holds.
\end{algorithmic}
\end{algorithm}

The result of this recipe is that the random linear map $L$ satisfies the RIP on the set $\SCal$ with constant $\delta \in (0, \underline{\delta}_p)$ and probability at least $1-\xi$ provided that \refeq{eq:cond_m_main_theorem} holds. 

In many practical scenarios, such as the ones discussed in next sections, the concentration inequalities \refeq{eq:prob_bound_increment_1} and \refeq{eq:prob_bound_increment_2} in Step $3$ are easy to obtain using well-known concentration inequalities for the sum of independent random variables, see, \eg, \cite{vershynin12, foucart13}. We will also see that the estimation of $\underline{\delta}_p$ simplifies in these cases. Therefore, the main difficult step in the recipe is most often the computation of the upper box-counting dimension of $\SCal$.

%
\section{A generic construction of $L$ with the RIP}
\label{sec:generic_linear_map}

In the previous section, we provided a recipe to prove that a random linear map satisfies the RIP once the probability distribution $\mu$ is given. In this section, we give a generic way of constructing a random linear map which has the RIP on a set $\SCal$ that satisfies Assumption \ref{as:dimension}, \ie, we give a generic construction of $\mu$. 

We divide our construction into two steps. In the first step, we build a continuous linear functional $b$ that maps the vectors in $\SCal$ to finite but potentially large dimension while essentially preserving their norm. In the second step, we further reduce the embedding dimension by multiplication with a random matrix.

\subsection{Mapping to a finite-dimensional subspace}
\label{sec:b_hilbert_space}

In this section, we prove that it is always possible to design a continuous linear map $b \colon \HCal \rightarrow  \Rbb^d$, where $d$ is potentially large but finite, such that
\begin{align}
\label{eq:upper_bound_proj}
\norm{b(\vec{x})}_{b} \leq \norm{\vec{x}},\ \forall \vec{x} \in \HCal,
\end{align}
and
\begin{align}
\label{eq:lower_bound_proj}
\norm{b(\vec{x})}_{b} \geq 1 - \epsilon_*, \ \forall \vec{x} \in \SCal,
\end{align}
for some $\epsilon_* \in (0, 1)$ and a well-chosen norm $\norm{\cdot}_{b}$. One can remark that the above properties ensure that $b$ already satisfies a RIP but for a potentially large dimension $d$. We will see how to further reduce the dimension in Section~\ref{sec:dim_red_random_matrix}.

\add{In the following subsections, we detail a generic method to construct $b$. This construction uses the fact that we can cover $\SCal$ with precision $\epsilon_*$ as Assumption~\ref{as:dimension} holds. Note that this construction can be computationally expensive in practice. Yet, we would like to highlight that one does not have to use such a covering in practice. Indeed, the only required property in the following results is that $b$ satisfies \eqref{eq:upper_bound_proj} and \eqref{eq:lower_bound_proj}. The way $b$ is constructed is not important as long as these properties hold. In practice, one can use other properties of $\SCal$ to construct $b$ efficiently. For example, we will see in Section~\ref{sec:mri_example} that, for signals sparse in the Haar wavelet basis of $L_2([0, 1])$, such a linear map can be built easily using properties of these wavelets in the Fourier basis.}

\subsubsection{Using an orthonormal basis of a well-chosen subspace}

\begin{figure}
\centering
\includegraphics[width=\columnwidth]{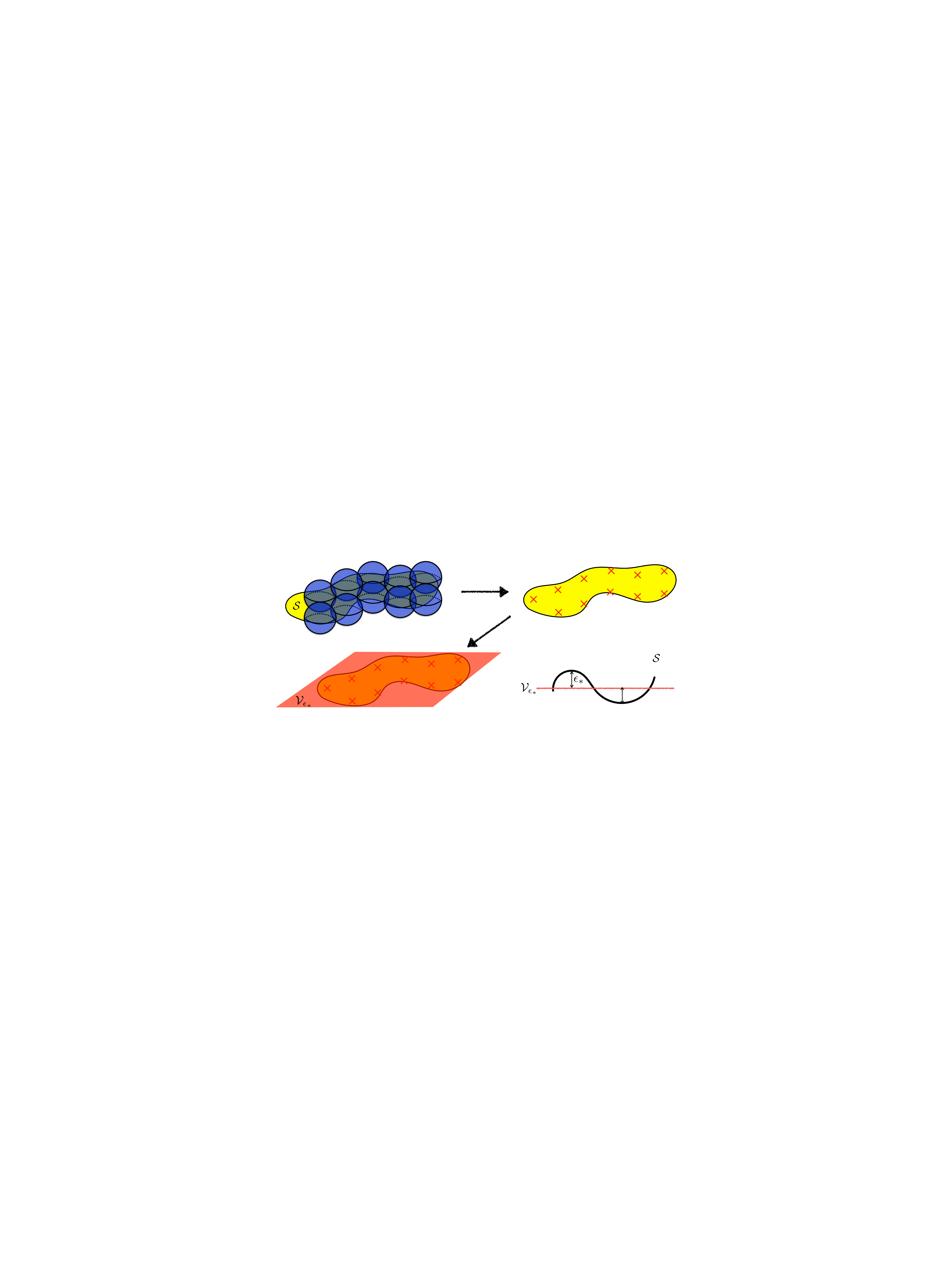}
\caption{\label{fig:contruction_V_epsilon_p} Construction of $\VCal_{\epsilon_*}$. \emph{Top left}: cover of $\SCal$ with $N(\epsilon_*)$ balls of radius $\epsilon_*$. \emph{Top right}: the centres of the balls, indicated by the red crosses, form a $\epsilon_*$-net, denoted by $\CCal(\epsilon_*)$, for $\SCal$. \emph{Bottom left}: the linear span of the vectors in $\CCal(\epsilon_*)$ is $\VCal_{\epsilon_*}$. \emph{Bottom right}: $\VCal_{\epsilon_*}$ approximates $\SCal$ with precision $\epsilon_*$.}
\end{figure}

If Assumption \ref{as:dimension} holds, then we can cover $\SCal$ with a finite number of closed balls. We fix a resolution $\epsilon_*$, find a minimum $\epsilon_*$-net for $\SCal$, and denote the set of points in this net by $\CCal(\epsilon_*)$. The cardinality of $\CCal(\epsilon_*)$ is $N_{\SCal}(\epsilon_*)$. Let $\VCal_{\epsilon_*} \subset \HCal$ be the finite-dimensional linear subspace of $\HCal$ spanned by $\CCal(\epsilon_*)$, and $P_{\VCal_{\epsilon_*}} : \HCal \rightarrow \HCal$ be the orthogonal projection onto $\VCal_{\epsilon_*}$. By construction of $\VCal_{\epsilon_*}$, we have
\begin{align}
\label{eq:bound_projection}
\sup_{\vec{x} \in \SCal} \norm{\vec{x} - P_{\VCal_{\epsilon_*}}(\vec{x})}
\leq
\sup_{\vec{x} \in \SCal} \min_{\vec{x}_0 \in \CCal(\epsilon_*)} \norm{\vec{x} - \vec{x}_0}
\leq
\epsilon_*. 
\end{align}
The orthogonal projection onto $\VCal_{\epsilon_*}$ thus preserves the norm of all vectors in $\SCal$ with an error at most $\epsilon_*$, as illustrated in Fig. \ref{fig:contruction_V_epsilon_p}.

Let $(\vec{b}_1, \ldots, \vec{b}_d)$ be an orthonormal basis for $\VCal_{\epsilon_*}$ and $b$ be the linear map 
\begin{align*}
b \colon \HCal 	& \longrightarrow  	\Rbb^d \nonumber \nonumber \\
 \vec{x} 					& \longmapsto 	(\scp{\vec{b}_i}{\vec{x}})_{1 \leq i \leq d}.
\end{align*}
We remark that 
\begin{align*}
\norm{b(\vec{x})}_2 = \norm{P_{\VCal_{\epsilon_*}}(\vec{x})} \leq \norm{\vec{x}},\ \forall \vec{x} \in \HCal,
\end{align*}
and that
\begin{align*}
\norm{b(\vec{x})}_2 \geq \norm{\vec{x}} - \norm{\vec{x} - P_{\VCal_{\epsilon_*}}(\vec{x})} \geq 1 - \epsilon_*, \ \forall \vec{x} \in \SCal.
\end{align*}
To obtain the last inequality, we used inequality~\refeq{eq:bound_projection} and the fact that $\SCal \subset \Sbb$. We thus proved the desired result with $\norm{\cdot}_{b} = \norm{\cdot}_2$.

%

\subsubsection{Using an arbitrary basis of $\VCal_{\epsilon_*}$}

The linear map $b$ in the last section was constructed using an orthonormal basis of $\VCal_{\epsilon_*}$. To provide more flexibility in the design of the linear map $L$, we propose here to use an arbitrary basis of $\VCal_{\epsilon_*}$. Let $(\vec{b}_1, \ldots, \vec{b}_d)$ be such a basis and define
\begin{align*}
b \colon \HCal 	& \longrightarrow 	\Rbb^d \\
\vec{x}		& \longmapsto 		(\scp{\vec{b}_i}{\vec{x}})_{1 \leq i \leq d}.
\end{align*}
We also define the following norm
\begin{align*}
\norm{y}_{b} = \inf_{\vec{z} \in \HCal} \left\{ \norm{\vec{z}} \ \vert \  b(\vec{z}) = y \right\},
\end{align*}
for all $y \in \Rbb^d$. We next show that \refeq{eq:upper_bound_proj} and \refeq{eq:lower_bound_proj} hold with this choice of norm.

We denote by $\vec{y} \in \VCal_{\epsilon_*}$ the vector with coordinates $y$ in the basis $(\vec{b}_1, \ldots, \vec{b}_d)$. Let $\VCal_{\epsilon_*}^{\bot}$ be the orthogonal complement to $\VCal_{\epsilon_*}$ in $\HCal$. It is easy to notice that the set of vectors that satisfy $b(\vec{z}) = y$ is
\begin{align*}
\{ \vec{z}~=~\vec{y} + \vec{y}^{\bot} \ \vert \ \vec{y}^{\bot} \in \VCal_{\epsilon^*}^{\bot} \},
\end{align*}
\ie, the set of vectors whose orthogonal projection on $\VCal_{\epsilon_*}$ is $\vec{y}$. Then, by orthogonality, we have $\norm{\vec{z}}^2 = \norm{\vec{y}}^2~+~\norm{\vec{y}^{\bot}}^2$, whose minimum value is $\norm{\vec{y}}^2$. Therefore, $\norm{y}_{b} = \norm{\vec{y}}$ and we deduce that $\norm{b(\vec{x})}_{b} = \norm{P_{\VCal_{\epsilon_*}}(\vec{x})}$ for all $\vec{x} \in \HCal$. As we have $\norm{P_{\VCal_{\epsilon_*}}(\vec{x})} \leq \norm{\vec{x}}$ for all $\vec{x} \in \HCal$, \refeq{eq:upper_bound_proj} holds. Then the fact that $\norm{P_{\VCal_{\epsilon_*}}(\vec{x})} \geq \norm{\vec{x}} - \norm{\vec{x} - P_{\VCal_{\epsilon_*}}(\vec{x})} \geq 1 - \epsilon_*$ for all $\vec{x} \in \SCal$ shows that \refeq{eq:lower_bound_proj} holds.

We remark that when $(\vec{b}_1, \ldots, \vec{b}_d)$ is an orthonormal basis, we have $\norm{y}_{b} = \norm{\vec{y}} = \norm{y}_{2}$, \ie, the norm we naturally chose in the previous section. The choice of $\norm{\cdot}_b$ as norm in $\Rbb^d$ allows us to recover the result of the last section but also to generalise it. Indeed, the result is now independent of the choice of the basis $(\vec{b}_1, \ldots, \vec{b}_d)$. As explained in the remark before Recipe $2$ below, this result can be useful when the choice of $(\vec{b}_1, \ldots, \vec{b}_d)$ is for example fixed by the application.

\subsection{Dimensionality reduction with a random matrix}
\label{sec:dim_red_random_matrix}

The linear map $b$ in the previous section maps the vectors in $\SCal$ into finite, but potentially large, dimension with an error at most $\epsilon_*$ on the norm of the vectors. The goal is now to reduce the dimension further without degrading much the norm of the vectors in $b(\SCal)$.

Before continuing, we introduce the following definitions (definitions $5.7$ and $5.13$ in \cite{vershynin12}).

\begin{definition}[Subexponential random variable]
\label{def:subexp}
A subexponential random variable $X$ is a random variable that satisfies
\begin{align*}
(\Ebb \abs{X}^q)^{1/q} \leq C q\ \text{for all}\ q \geq 1,
\end{align*}
with $C>0$. The subexponential norm of $X$, denoted by $\norm{X}_{\Psi_1}$, is the smallest constant $C$ for which the last property holds, \ie,
\begin{align*}
\norm{X}_{\Psi_1} := \sup_{q\geq1} \left\{ q^{-1} \; (\Ebb \abs{X}^q)^{1/q} \right\}.
\end{align*}
\end{definition}
\begin{definition}[Subgaussian random variable]
\label{def:subgauss}
A subgaussian random variable $X$ is a random variable that satisfies 
\begin{align*}
(\Ebb \abs{X}^q)^{1/q} \leq C \sqrt{q}\ \text{for all}\ q \geq 1,
\end{align*}
with $C>0$. The subgaussian norm of $X$, denoted by $\norm{X}_{\Psi_2}$, is the smallest constant $C$ for which the last property holds, \ie,
\begin{align*}
\norm{X}_{\Psi_2} := \sup_{q\geq1} \left\{ q^{-1/2} \; (\Ebb \abs{X}^q)^{1/q} \right\}.
\end{align*}
\end{definition}

To reduce the dimension, we draw $m$ independent random vectors $a_1, \ldots, a_m \in \Rbb^d$ according to a probability distribution $\nu$ in $\Rbb^d$. The measurements $a_i^\adjoint b(\vec{x})$, $i = 1, \ldots, m$ are thus independent and identically distributed random variables. The choice of $b$ and $\nu$ defines the probability distribution $\mu$ in $\HCal^m$. 

To characterise the number of measurements needed to preserve the norm of all vectors in $\SCal$, we need the following quantities
\begin{align}
\Lambda_1 & := \sup_{\substack{\vec{x} \in \{ \SCal - \SCal\} \cup  \SCal \\ \vec{x} \neq \vec{0} }} \left\{ {\norm{a_i^\adjoint b(\vec{x})}_{\Psi_1}}/{\norm{b(\vec{x})}_{b}} \right\}\ \text{and}\\
\Lambda_2 & := \sup_{\substack{\vec{x} \in \{ \SCal - \SCal\} \cup  \SCal \\ \vec{x} \neq \vec{0} }} \left\{ {\norm{a_i^\adjoint b(\vec{x})}_{\Psi_2}}/{\norm{b(\vec{x})}_{b}} \right\}.
\end{align}
Note that these quantities may be infinite. However, we can ensure that $\Lambda_1$ and $\Lambda_2$ are finite, for example, by drawing $a_1, \ldots, a_m$ in the ball of radius $1$ with respect to the dual norm of $\norm{\cdot}_{b}$ defined as
\begin{align*}
\norm{a}_{b^*} :=  \sup \left\{ \abs{a^\adjoint x} \, \vert \, x \in \Rbb^d, \norm{x}_{b} \leq 1 \right\}.
\end{align*}
Indeed, we have $\abs{a^\adjoint b(\vec{x})} \leq \norm{b(\vec{x})}_{b}$ for any $\vec{x} \in \HCal$ in this case, which ensures that $\Lambda_2 \leq 1$ and $\Lambda_1 \leq 1$. 

We are now ready to give the two main results of this section, which follow from Theorem \ref{th:main_theorem}.

\begin{theorem} 
\label{th:rip_1}
Assume that $\Lambda_1 < +\infty$ and define
\begin{align}
\label{eq:L_p=1}
L \colon \HCal 	& \longrightarrow  	\Rbb^m \nonumber \nonumber \\
\vec{x} 		& \longmapsto 		\left(\frac{a_i^\adjoint b(\vec{x})}{m}\right)_{1 \leq i \leq d}.
\end{align}
There exists an absolute constant $C > 0$ such that if Assumption~\ref{as:dimension} holds then, for any $\xi \in (0, 1)$ and $\delta \in (0, 1)$, with probability at least $1 - \xi$,
\begin{align*}
\norm{\vec{x}}_{\mu, 1} - \delta \leq \norm{L(\vec{x})}_1 \leq  \norm{\vec{x}}_{\mu, 1} + \delta,
\end{align*}
holds for all $\vec{x} \in \SCal$ provided that
\begin{align}
\label{eq:m_p=1}
m \geq \frac{C}{\delta^2} \max(2\Lambda_1^2, \Lambda_1) \max\left\{s \log\left(\inv{\epsilon_{\SCal}}\right),\ \log\left(\frac{6}{\xi}\right) \right\}.
\end{align}
\end{theorem}
\begin{theorem} 
\label{th:rip_2}
Assume that $\Lambda_2 < +\infty$ and define
\begin{align}
\label{eq:L_p=2}
L \colon \HCal 	& \longrightarrow  	\Rbb^m \nonumber \nonumber \\
\vec{x} 		& \longmapsto 	\left(\frac{a_i^\adjoint b(\vec{x})}{\sqrt{m}}\right)_{1 \leq i \leq d}.
\end{align}
There exists an absolute constant $C > 0$ such that if Assumption~\ref{as:dimension} holds, then, for any $\xi \in (0, 1)$ and $\delta \in (0, 1)$, with probability at least $1 - \xi$,
\begin{align*}
\norm{\vec{x}}_{\mu, 2}^2 - \delta \leq \norm{L(\vec{x})}_2^2 \leq \norm{\vec{x}}_{\mu, 2}^2 + \delta,
\end{align*}
holds for all $\vec{x} \in \SCal$ provided that
\begin{align}
\label{eq:m_p=2}
m \geq \frac{C}{\delta^2} \max(8\Lambda_2^4, \Lambda_2^2) \max\left\{s \log\left(\inv{\epsilon_{\SCal}}\right),\ \log\left(\frac{6}{\xi}\right) \right\}.
\end{align}
\end{theorem}

Both theorems are proved in Appendix \ref{app:proof_theorem_rip_1_2}. Let us comment on these results and highlight the situations where they are useful. We will study several examples in the next section.

\begin{itemize} 
\item First, we remark that 
\begin{align*}
\underline{\delta}_p = \inf_{\vec{x} \in \SCal} \Ebb_{\nu} \abs{a_i^\adjoint b(\vec{x})}^p \ \leq \ \sup_{\vec{x} \in \SCal} \Ebb_{\nu} \abs{a_i^\adjoint b(\vec{x})}^p = \bar{\delta}_p
\end{align*}
with $p=1$ in Theorem \ref{th:rip_1} and $p=2$ in Theorem \ref{th:rip_2}. By definition of $\Lambda_1$ and $\Lambda_2$, we have
\begin{align*}
\Ebb_{\nu} \abs{a_i^\adjoint b(\vec{x})}^p 
\; \leq \; p \, \Lambda_p^p \, \norm{b(\vec{x})}_{b}^p
\; \leq \; p \, \Lambda_p^p,\quad \forall \vec{x} \in \SCal,
\end{align*}
for $p = 1$ and $2$. The last inequality follows from \refeq{eq:upper_bound_proj}. Therefore,
\begin{align*}
\underline{\delta}_p \; \leq \; \bar{\delta}_p \; \leq \; p \, \Lambda_p^p
\end{align*}
with $p=1$ in Theorem \ref{th:rip_1} and $p=2$ in Theorem \ref{th:rip_2}.
\item Second, we notice that the RIP is satisfied for $\delta < \underline{\delta}_1$ in the case of Theorem \ref{th:rip_1} and for $\delta < \underline{\delta}_2$ in the case of Theorem \ref{th:rip_2}. Consequently, if $L$ is constructed as in this section, then Recipe \ref{alg:recipe_rip} can be modified to Recipe \ref{alg:recipe_rip_bis}.
\begin{algorithm}[h]
\caption{Recipe to prove that $L$ in \refeq{eq:L_p=1} or \refeq{eq:L_p=2} satisfies the RIP on $\SCal$}
\label{alg:recipe_rip_bis}
\begin{algorithmic}[1]
\State Prove that the set $\SCal$ has finite upper box-counting dimension (Assumption \ref{as:dimension}).
\State Compute $\underline{\delta}_p = \inf_{\vec{x} \in \SCal} \Ebb_{\nu} \abs{a_i^\adjoint b(\vec{x})}^p$, with $p=1$ for $L$ defined in \refeq{eq:L_p=1} or $p=2$ for $L$ defined in \refeq{eq:L_p=2}.
\State Compute $\Lambda_1$ or $\Lambda_2$.
\State Choose $m$ such that \refeq{eq:m_p=1} holds if $p=1$, or such that \refeq{eq:m_p=2} holds if $p=2$.
\end{algorithmic}
\end{algorithm}
\item Third, the number of measurements essentially scales with $s \, \max(2\Lambda_1^2, \Lambda_1)/\delta^2$ in the first case and $s \, \max(8\Lambda_2^4, \Lambda_2^2)/\delta^2$ in the second case. As we should have $\delta < \underline{\delta}_p$ (with $p = 1$ or $p = 2$) to satisfy the RIP, the results have a practical interest when $\max(2\Lambda_1^2, \Lambda_1)/\underline{\delta}_1^2$ or $\max(8\Lambda_2^4, \Lambda_2^2)/\underline{\delta}_2^2$ is small. These ratios are the quantities to optimise when designing $L$, as illustrated in Section \ref{sec:rop}.
\item Finally, it is important to notice that the choice of the basis $(\vec{b}_1, \ldots, \vec{b}_d)$ and the choice of the distribution $\nu$ for the $a_i$'s interact together in the value of the ratio $\max(2\Lambda_1^2, \Lambda_1)/\underline{\delta}_1^2$ or $\max(8\Lambda_2^4, \Lambda_2^2)/\underline{\delta}_2^2$. If one has the flexibility to choose both $(\vec{b}_1, \ldots, \vec{b}_d)$ and $\nu$, then one should seek to minimise these ratios in order to minimise $m$. Even if the choice of $b$ is fixed by the application of interest, one still has the flexibility to optimise the distribution $\nu$ to minimise $m$.
\end{itemize}

In Recipe \ref{alg:recipe_rip_bis} \add{introduced} above, when $(\vec{b}_1, \ldots, \vec{b}_d)$ is an orthonormal basis and $a_i$ is an isotropic random vector, then $\underline{\delta}_2$ can be directly estimated using \refeq{eq:bound_b_hilbert} as done below in Section \ref{sec:rip_hilbert_infinite} for the second step of the recipe. We recall that $a_i \in \Rbb^d$ is isotropic if $\Ebb\abs{a_i^\adjoint x}^2 = \norm{x}_2^2$ for all $x \in \Rbb^d$. To estimate $\underline{\delta}_1$, one can use Lemma \ref{lemma:subexp_bound_expected} as also done below in Section \ref{sec:RIP_ROP_gauss} and Section \ref{sec:RIP_ROP_sparse} for the second step of the recipe.

\section{Examples}
\label{sec:examples}

In this section, we show how to use our generic recipe on different examples.

\subsection{A linear embedding from $\HCal$ to $\Rbb^m$}
\label{sec:rip_hilbert}

\subsubsection{The infinite-dimensional case}
\label{sec:rip_hilbert_infinite}

In Section \ref{sec:b_hilbert_space}, we built a linear map $b : \HCal \rightarrow \Rbb^d$ which preserves the norm of all vectors in $\SCal$. For simplicity, we consider the case where $(\vec{b}_1, \ldots, \vec{b}_d)$ is an orthonormal basis. We have
\begin{align}
\label{eq:bound_b_hilbert}
(1 - \epsilon_*)^2 \norm{\vec{x}}^2 \leq \norm{b(\vec{x})}_b^2 = \norm{b(\vec{x})}_2^2 \leq \norm{\vec{x}}^2, \quad \forall \vec{x} \in \SCal.
\end{align}
Let $a_1, \ldots, a_m \in \Rbb^d$ be $m$ vectors whose entries are independent Gaussian random variables with mean $0$ and variance $1$, and define $L$ as in Theorem \ref{th:rip_2}. We follow Recipe \ref{alg:recipe_rip_bis} to prove that $L$ has the RIP.

\begin{enumerate}[1:]
\item The set $\SCal$ has a finite upper-box counting dimension by assumption.
\item To estimate $\underline{\delta}_2$, we first notice that
\begin{align*}
\Ebb \abs{a_i^\adjoint b(\vec{x})}^2 = \norm{b(\vec{x})}_2^2.
\end{align*}
Inequality \refeq{eq:bound_b_hilbert} then yields
\begin{align*}
(1-\epsilon_*)^2 \norm{\vec{x}}^2 \leq \Ebb \abs{a_i^\adjoint b(\vec{x})}^2 \leq \norm{\vec{x}}^2, \quad \forall \vec{x} \in \SCal.
\end{align*}
Therefore, $\underline{\delta}_2 \geq (1-\epsilon_*)^2$ and $\bar{\delta}_2 \leq 1$. 
\item To estimate $\Lambda_2$, we use the fact that $\norm{a_i^\adjoint x}_{\Psi_2} \leq D \norm{x}_2$, for all $x \in \Rbb^d$, where $D>0$ is an absolute constant (see \cite{vershynin12}). Therefore, 
\begin{align*}
\norm{a_i^\adjoint b(\vec{x})}_{\Psi_2} \leq D \norm{b(\vec{x})}_2, \quad \forall \vec{x} \in \HCal,
\end{align*}
and $\Lambda_2 \leq D$. 
\item The last step of the recipe proves that $L$ satisfies the RIP with constant $\delta \in (0, (1-\epsilon_*)^2)$ and probability at least $1-\xi$, \ie, 
\begin{align*}
(1-\epsilon_*)^2 - \delta \leq \norm{L(\vec{x})}_2^2  \leq  1+ \delta, \quad \forall \vec{x} \in \SCal,
\end{align*}
provided that 
\begin{align*}
m \geq \frac{C}{\delta^2} \max\left\{s \log\left(\inv{\epsilon_{\SCal}}\right),\ \log\left(\frac{6}{\xi}\right) \right\},
\end{align*}
where $C$ is an absolute constant. 
\end{enumerate}
With this example, one can remark that the only difference in the construction of the linear map $L$ between an infinite-dimensional setting and a finite-dimensional setting is the presence of the intermediate mapping $b$. This mapping is built from the projection onto a well-chosen subspace that preserves the norm of all vectors in $\SCal$. Finally, the number of measurements $m$ is essentially proportional to the intrinsic dimension of $\SCal$, as usual in CS results.

\subsubsection{Recovery of known results in finite dimension}
\label{sec:known_result_hilbert}

The above results holds for any set $\SCal$ that satisfies Assumption~\ref{as:dimension} in a Hilbert space $\HCal$, so it also holds for signal models in $\Rbb^n$ provided that they have a finite upper box-counting dimension. Let us take one example in finite ambient space. 

Consider the set $\SCal$ of $2k$-sparse signals with unit $\ell_2$-norm in $\Rbb^n$. This set can be covered by at most $[3\ee n/(2k\epsilon)]^{2k}$ balls of radius $\epsilon \in (0, 1)$ \cite{foucart13}. Its upper box-counting dimension is thus $2k$. Let us compute $\epsilon_{\SCal}$ and $s$ which appear in Assumption \ref{as:dimension}. We remind that $\epsilon_{\SCal}$ is a constant such that $[3\ee n/(2k\epsilon)]^{2k} \leq \epsilon^{-s}$ for all $\epsilon \leq \epsilon_{\SCal}$ with $s > 2k$. Writing $s = 2k + \eta$ with $\eta>0$, we should have $(3\ee n/(2k))^{2k} \leq \epsilon^{-\eta}$ for all $\epsilon \leq \epsilon_{\SCal}$, or, equivalently, $(3\ee n/(2k))^{2k} \leq \epsilon_{\SCal}^{-\eta}$. We take $\epsilon_{\SCal} = 2k/(3\ee n)$ and $\eta = 2k$. Notice that $\epsilon_{\SCal} \leq 1/2$ and that $s = 4k$. As we are in finite ambient dimension, we can take $\VCal_{\epsilon_*} = \Rbb^n$ (which implies that $\epsilon_* = 0$ and $d=n$) and the canonical basis for $(\vec{b}_1, \ldots, \vec{b}_n)$ (which implies that $b$ is the identity). The complete linear map $L$ thus reduces to the matrix $\ma{A}/\sqrt{m}$ which satisfies, with probability at least $1-\xi$,
\begin{align}
\label{eq:usual_RIP_sparse}
(1 - \delta) \norm{\vec{x}}_2^2 \leq \frac{\norm{\ma{A}\vec{x}}_2^2}{m} \leq (1 + \delta) \norm{\vec{x}}_2^2,
\end{align}
for all $2k$-sparse vectors $\vec{x}$ provided that $m$ satisfies
\begin{align*}
m \geq \frac{C}{\delta^2} \max\left\{4k \log\left(\frac{3\ee n}{2k}\right) ,\ \log\left(\frac{6}{\xi}\right) \right\}.
\end{align*}
In particular, this result shows that the set of $k$-sparse vectors is stably embedded into $\Rbb^m$ for $m$ satisfying the above inequality. 

In comparison, Theorem $9.2$ in \cite{foucart13} shows that $\ma{A}/\sqrt{m}$ satisfies~\refeq{eq:usual_RIP_sparse} with probability at least $1 - \xi$ provided that
\begin{align*}
m \geq \frac{D}{\delta^2} \; \left(2k \log\left(\frac{\ee n}{2k}\right) +  \log\left(\frac{2}{\xi}\right)\right),
\end{align*}
where $D>0$ is an absolute constant. This condition on $m$ is similar to ours. This shows that we recover results similar to known ones in CS.

Note that we could have chosen any orthonormal basis of $\Rbb^n$ for $(\vec{b}_1, \ldots, \vec{b}_n)$. This illustrates the known universality of subgaussian measurement matrices relative to the sparsity basis \cite{foucart13}.

Following the same procedure, one can easily recover many other similar results for, \eg, the set of low-rank matrices \cite{candes11a}, low-dimensional smooth manifolds \cite{eftekhari14}, or the set of group-sparse signals \cite{ayaz14}. In all these cases, the only remaining difficulty is estimating the upper box-counting dimension of the set of interest.

\subsection{Embedding of matrices with rank-one projections}
\label{sec:rop}

As a second example, we propose to use our results to show that one can embed a low dimensional set of matrices using rank-one projections. \add{This scheme was proposed and studied in, \eg, \cite{cai15, chen15, kueng15, bahmani15} to embed certain low-dimensional set of matrices: low-rank matrices, sparse matrices, low-rank+sparse matrices, \emph{etc}}. We extend here these results to any set with a low-dimensional normalised secant set that satisfies Assumption \ref{as:dimension}, which, \eg, includes matrix manifolds.

In this section, the ambient space is the space of real matrices of size $n_1 \times n_2$, $\Rbb^{n_1 \times n_2}$, equipped with the Frobenius norm, $\norm{\cdot}_{\rm Frob}$. We consider that $\SCal \subset \Rbb^{n_1 \times n_2}$ is a low-dimensional subset of matrices that satisfies Assumption \ref{as:dimension}, with $\norm{\cdot}_{\rm Frob}$ instead of $\norm{\cdot}$.

\subsubsection{Measurement strategy}

As done in \cite{cai15, chen15, kueng15, bahmani15}, we propose to measure a matrix $\ma{M} \in \Rbb^{n_1 \times n_2}$ using $m$ rank-one projections:
\begin{align}
\label{eq:ROP_linear_map}
L \colon \Rbb^{n_1 \times n_2} 	 &  \longrightarrow   	\Rbb^m \nonumber \\
 \ma{M} 		 & \longmapsto  \left(\frac{a_i^\adjoint \ma{M} b_i}{m}\right)_{i=1, \ldots, m},
\end{align}
where $a_1, \ldots, a_m \in \Rbb^{n_1}$ and $b_1, \ldots, b_m \in \Rbb^{n_2}$ are independent random vectors. Note that the main advantage of this measurement strategy is its memory efficiency. Indeed, we only need to store $m \, (n_1+n_2)$ coefficients to compute the measurements while the measurement strategy proposed in \cite{candes11a} requires to store $m \, n_1n_2$ coefficients.

We will pay attention to two types of random vectors. First, we will focus on the case where the entries of the vectors $a_i$ and $b_i$, $i = 1, \ldots, m$, are independent draws of a random variable $N \in \Rbb$ that has a standard normal distribution. Second, we will discuss the case where the entries of the vectors $a_i$ and $b_i$, $i = 1, \ldots, m$, are independent draws of a random variable $P \in \{ 0,  \pm \sqrt{q} \}$ that satisfies
\begin{align}
\label{eq:sparse_random_variable}
\Prob\,(P = 0) = \frac{q-1}{q}\ \text{ and }\ \Prob\, \left(P = \pm \sqrt{q} \right) = \frac{1}{2q},
\end{align}
where $q \geq 1$. The larger $q$ is, the sparser (on average) the measurement vectors are, hence improving the computational efficiency of the measurement strategy. In this case, the average sparsity of the vectors $a_i$ and $b_i$ is $n_1/q$ and $n_2/q$, respectively, and computing one measurement can be done in $O(n_1n_2/q^2)$ operations. This type of measurement strategy with sparse vectors was proposed in \cite{achlioptas03} for Johnson-Lindenstrauss embeddings and can also be used for compressive sensing \cite{baraniuk08}. We will see that the number of measurements needed to preserve the norm of all vectors in $\SCal$ depends on the parameter $q$.

\subsubsection{RIP for rank-one projections with gaussian vectors}
\label{sec:RIP_ROP_gauss}

In this section, we study the case where the entries of the vectors $a_i$ and $b_i$ are independent draws of the random variable $N$. As the linear map $L$ has the form presented in Theorem \ref{th:rip_1} (with $b$ the identity and $\norm{\cdot}_b = \norm{\cdot}_{\rm Frob}$), we can follow Recipe \ref{alg:recipe_rip_bis} to prove that $L$ has the RIP.

\begin{enumerate}[1:]
\item The set $\SCal$ has a finite upper-box counting dimension by assumption.
\item To estimate $\underline{\delta}_1$, we use the result presented after Lemma~\ref{lemma:subexp_bound_expected} which shows that there exists an absolute constant $D>0$ such that
\begin{align*}
\underline{\delta}_1 := \inf_{\ma{M} \in \SCal} \Ebb\, \abs{a_i^\adjoint \ma{M} b_i} \geq D.
\end{align*}
Then, using the fact that
\begin{align*}
\Ebb \abs{a_i^\adjoint \ma{M} b_i} \leq \left( \Ebb \abs{a_i^\adjoint \ma{M} b_i}^2 \right)^{1/2} = \norm{\ma{M}}_{\rm Frob},\quad \forall \, \ma{M} \in \Rbb^{n_1 \times n_2},
\end{align*}
we also deduce that $\bar{\delta}_1 \leq 1$. 
\item To estimate $\Lambda_1$, we use Lemma \ref{lemma:subexp_ROP} which shows that there exists an absolute constant $E>0$ such that
\begin{align*}
\norm{a_i^\adjoint \ma{M} b_i}_{\Psi_1} \leq E \norm{\ma{M}}_{\rm Frob} \quad \forall \, \ma{M} \in \Rbb^{n_1 \times n_2}.
\end{align*}
Therefore $\Lambda_1 \leq E$. 
\item The last step of the recipe proves that L satisfies the RIP with constant $\delta \in (0, D)$ and probability at least $1-\xi$, \ie, 
\begin{align*}
D - \delta \leq \norm{L(\ma{M})}_1 \leq 1 + \delta \quad \forall \, \ma{M} \in \SCal,
\end{align*}
provided that 
\begin{align*}
m \geq \frac{C}{\delta^2} \max\left\{s \log\left(\inv{\epsilon_{\SCal}}\right),\ \log\left(\frac{6}{\xi}\right) \right\},
\end{align*}
where $C>0$ is an absolute constant. The number of measurements $m$ thus scales linearly with $s$. This result is true for \emph{any set $\SCal$ that satisfies Assumption \ref{as:dimension}} and is not restricted to the set of low-rank matrices.
\end{enumerate}

Let us highlight that the result presented in this section involves an $\ell_1$-norm in the measurement domain, instead of the $\ell_2$-norm as in most CS results. The authors of \cite{cai15} and \cite{chen15} also chose the $\ell_1$-norm in the measurement domain to prove the RIP. In addition, the authors in \cite{cai15} proved that if one chooses the $\ell_2$-norm in the measurement domain for the RIP, then at least $O(n_1n_2)$ measurements are needed to ensure that $L$ satisfies a RIP condition that guarantees the recovery of rank-$1$ matrices (see Lemma $2.1$ in \cite{cai15} and the discussion thereafter), which is much larger than $n_1 + n_2 -1$, the number of degrees of freedom for rank-$1$ matrices. Choosing the $\ell_1$-norm in the measurement domain solves this issue.

\subsubsection{Recovery of a known result}

Let now $\SCal$ be the set of rank-$2r$ matrices with unit Frobenius norm. Lemma $3.1$ in \cite{candes11a} shows that $N_{\SCal} (\epsilon) \leq (9/\epsilon)^{2r(n_1 + n_2 + 1)}$ for any $\epsilon>0$. The upper box-counting dimension of $\SCal$ is thus $2r(n_1 + n_2 + 1)$. Using the procedure of Section \ref{sec:known_result_hilbert}, we can satisfy Assumption \ref{as:dimension} by taking $\epsilon_{\SCal} = 1/9$ and $s = 4r(n_1 + n_2 + 1)$. The above result then shows that for $\delta \in (0, E)$, with probability at least $1-\xi$, 
\begin{align*}
E - \delta \leq \norm{L(\ma{M})}_1 \leq 1 + \delta, \quad \forall \, \ma{M} \in \SCal
\end{align*}
provided that 
\begin{align*}
m \geq \frac{C}{\delta^2} \max\left\{2r(n_1 + n_2 + 1) \log\left(9\right),\ \log\left(\frac{6}{\xi}\right) \right\},
\end{align*}
where $C>0$ is an absolute constant. In particular, the set of rank-$r$ matrices is stably embedded into $\Rbb^m$ for $m$ satisfying the above inequality.

Let us compare this result with Theorem $2.2$ in \cite{cai15}.
\begin{theorem}[Theorem 2.2, \cite{cai15}] 
Let $L : \Rbb^{n_1 \times n_2} \rightarrow \Rbb^m$ be the linear map defined in \refeq{eq:ROP_linear_map}. For positive numbers\footnote{These numbers depend on the subgaussian norm of the measurement vectors $a_i$ and $b_i$.} $C_1 < 1/3$ and $C_2>1$, there exist constants $C$ and $\delta$, not depending on $n_1$, $n_2$, and $r$, such that if $m \geq C r (n_1+n_2)$, then with probability at least $1 - \ee^{-m\delta}$, $L$ satisfies 
\begin{align*}
C_1 \norm{\ma{M}}_{\rm Frob} \leq \norm{L(\ma{M})}_1 \leq C_2 \norm{\ma{M}}_{\rm Frob},
\end{align*}
for all rank-$2r$ matrices $\ma{M} \in \Rbb^{n_1 \times n_2}$. 
\end{theorem}

Comparing our result and the theorem above, we notice that we recover the same type of result with a number of measurements that needs to be essentially proportional to the upper box-counting dimension of the set of rank-$r$ matrices in order to guarantee that the Frobenius norm of these matrices is preserved.

\subsubsection{RIP for rank-one projections with sparse vectors}
\label{sec:RIP_ROP_sparse}

To improve the computational efficiency of the measurement procedure, one can think of using sparse vectors $a_i$ and $b_i$. Let us thus discuss the case where the entries of the vectors $a_i$ and $b_i$ are independent draws of the sparse random variable $P$ according to~\eqref{eq:sparse_random_variable}. Recipe \ref{alg:recipe_rip_bis} still works to prove that $L$ has the RIP. 

\begin{enumerate}[1:]
\item The set $\SCal$ has a finite upper-box counting dimension by assumption.
\item To estimate $\underline{\delta}_1$, we use the result presented after Lemma~\ref{lemma:subexp_bound_expected} which shows that there exists an absolute constant $D>0$ such that
\begin{align*}
\underline{\delta}_1 := \inf_{\ma{M} \in \SCal} \Ebb\, \abs{a_i^\adjoint \ma{M} b_i} \geq \frac{D}{q\,(1+\log q)},
\end{align*}
for $q\geq2$. We also have $\bar{\delta}_1 \leq 1$, as with Gaussian random measurement vectors.
\item To estimate $\Lambda_1$, we use Lemma \ref{lemma:subexp_ROP} which shows that $\Lambda_1 \leq E \, q$, where $E>0$ is an absolute contant.  
\item The last step of Recipe \ref{alg:recipe_rip_bis} proves that $L$ satisfies the RIP with constant $\delta \in (0, D/(q+q\log q))$ and probability at least $1-\xi$, \ie, 
\begin{align*}
\frac{D}{q} - \delta \leq \norm{L(\ma{M})}_1 \leq 1 + \delta \quad \forall \, \ma{M} \in \SCal,
\end{align*}
provided that 
\begin{align*}
m \geq \frac{C \, q^2}{\delta^2} \max\left\{s \log\left(\inv{\epsilon_{\SCal}}\right),\ \log\left(\frac{6}{\xi}\right) \right\},
\end{align*}
where $C>0$ is an absolute constant.
\end{enumerate}

The number of measurements $m$ thus still scales linearly with $s$ but also with $q^{4}(1+\log q)^2$, when $\delta < D/(q+q\log q)$. As the (average) cost of computing one measurement is $O(n_1n_2/q^2)$, this result is too weak in general to prove any benefit of using sparse measurement vectors.

\subsubsection{\add{Discussion}}
Let us highlight that the above result is derived without using any properties of $\SCal$ beyond its box counting dimension. In particular, this result also applies for sets $\SCal$ of sparse matrices for which we guess that using very sparse measurement vectors is not an adequate measurement strategy. \add{Indeed, we need the support of the measurement vectors and the support of the vectors in the sparse matrix to both overlap in order to have non-zero measurements. As the measurement vectors get sparser, we increase the chance of having only null measurements, and thus decrease the chance of embedding the original matrix. This fact is reminiscent of the fundamental incoherence property between the sparsity domain and the sampling domain required in compressed sensing: one cannot sample a signal in the domain where it is sparse.} More optimistic results might be obtained by exploiting additional structures in $\SCal$. In that perspective, the work \cite{bourgain15} on sparse dimensionality reduction in Euclidean space might be particularly useful to obtain more optimistic results. The authors of \cite{bourgain15} prove that, with high probability, some random sparse measurement matrices satisfy the RIP for a wide class of subsets of the unit sphere. They provide sufficient conditions on the number of measurements and on the sparsity of the measurement matrix to ensure that the RIP is satisfied. These bounds depend on additional properties of $\SCal$ beyond its box-counting dimension. For example, for sparse signals in an orthonormal basis, they further require an incoherence property to ensure that the RIP holds (Section $5.1$, \cite{bourgain15}).

This example illustrates the importance of the ratio $\max(2\Lambda_1^2, \Lambda_1)/\underline{\delta}_1^2$ in the design of the linear map $L$. Even though the measurement vectors $a_i, b_i$ are subgaussian random vectors as in Section \ref{sec:RIP_ROP_gauss}, the ratio $\max(2\Lambda_1^2, \Lambda_1)/\underline{\delta}_1^2$ is too large and the result does not predict any gain in the computational efficiency compared to Gaussian rank-one measurements.

\section{Related works}
\label{sec:related_works}

\subsection{Infinite dimensional CS and generalised sampling}
\label{sec:mri_example}

An extension of the CS theory to an infinite dimensional setting is also studied in \cite{adcock14}. In their work, the ambient space is a separable Hilbert space $\HCal$ and the authors concentrate on the particular case of sparse signals in an orthonormal basis of $\HCal$. Their result does not involve a RIP but the linear map used to sense sparse signals shares some similarities with our construction of $L$. 

Let us take the example of $s$-sparse signals in an orthonormal basis $(\vec{\psi}_j)_{j \in \Nbb}$ of $\HCal$, \eg, one can think of a wavelet basis. For simplicity, let us consider that the support of the signals in $\SCal$ belongs to $\{0, \ldots,  n-1\}$, hence $\SCal$ has a finite upper box-counting dimension. In Section \ref{sec:rip_hilbert}, we measure the signals by first projecting them onto a finite dimensional subspace of large dimension $d$ that approximates the set $\SCal$. To construct this subspace, we propose here to use another orthonormal basis $(\vec{\phi}_i)_{i \in \Nbb}$ of $\HCal$, \eg, one can think of the Fourier basis. We choose to construct this finite dimensional subspace by using only the first $d$ basis vectors $\vec{\phi}_i$. Let us define the matrix $\ma{U} = (u_{ij} := \scp{\vec{\phi}_i}{\vec{\psi}_j})_{i, j \in \Nbb}$. Let $\alpha$ denote the coordinates of $\vec{x} \in \HCal$ in $(\vec{\psi}_j)_{j \in \Nbb}$. We have $b(\vec{x}) = \ma{P}_d \ma{U} \alpha \in \Rbb^d$, where $\ma{P}_d$ is the matrix that selects the first $d$ rows of $\ma{U}$. According to our recipe, we need to choose $d$ such that
\begin{align*}
(1-\epsilon_*) \norm{\vec{x}} \leq \norm{b(\vec{x})}_{2} \leq \norm{\vec{x}},\ \forall \vec{x} \in \SCal,
\end{align*}
in order to have the RIP. Here, this condition is equivalent to
\begin{align}
\label{eq:projection_condition}
(1-\epsilon_*) \norm{\ma{U} \ma{P}_n \alpha}_2 \leq \norm{\ma{P}_d \ma{U} \ma{P}_n \alpha}_{2} \leq \norm{\ma{U} \ma{P}_n \alpha}_2,
\end{align}
for all $\ma{U} \ma{P}_n \alpha \in \SCal \subset \Sbb$, where $\ma{P}_n$ is the matrix that keeps the first $n$ entries of $\alpha$ and set all the other ones to $0$. Using the facts that $\norm{\ma{U} \ma{P}_n \alpha}_2 = \norm{\ma{P}_n \alpha}_2$ and $\ma{P}_n^\adjoint \ma{P}_n = \ma{P}_n \ma{P}_n = \ma{P}_n$, taking the square of \refeq{eq:projection_condition} and rearranging the terms, one can notice that the condition
\begin{align}
\label{eq:balancing_property}
\norm{\ma{P}_n \ma{U}^\adjoint \ma{P}_d^\adjoint \ma{P}_d \ma{U} \ma{P}_n - \ma{P}_n}_{2} \leq \epsilon_*,
\end{align}
is sufficient to ensure \refeq{eq:projection_condition}. 

It is interesting to notice that this condition also appears in the generalised sampling theorem of \cite{adcock12}. Indeed, when $\ma{U}$ is an isometry as in the present example, the condition for stable reconstruction presented in \cite{adcock12} amounts to the control of $\norm{\ma{P}_n \ma{U}^\adjoint \ma{P}_d^\adjoint \ma{P}_d \ma{U} \ma{P}_n - \ma{P}_n}_{2}$ (see Section $5$ in \cite{adcock12} and the proofs therein). It is also proved in \cite{adcock12} that this quantity tends to $0$ as $d$ tends to infinity. This shows that to choose $d$ in practice, one just needs to increase it until \refeq{eq:balancing_property} is satisfied. In the particular case where $\HCal = L_2([0, 1])$, $(\vec{\psi}_j)_{j \in \Nbb}$ is the Haar wavelet basis (ordered from coarse to fine scale), and $(\vec{\phi}_j)_{j \in \Nbb}$ is the Fourier basis (ordered from low to high frequencies), the authors of \cite{adcock12} further prove that \refeq{eq:balancing_property} is satisfied for $d = O(n^2)$, and their numerical experiments suggest that it is already sufficient to take $d = O(n)$ (see Section $5.4$ in \cite{adcock12}). This indicates that $\VCal_{\epsilon_*}$ does not always need to be a very large dimensional subspace of $\HCal$ to ensure that the RIP holds.

Finally, we note that condition \refeq{eq:balancing_property} also appears in the extension of the CS theory to an infinite dimensional setting presented in \cite{adcock14}, involving however another operator norm. This property is called the weak balancing property in \cite{adcock14}.

\subsection{Other measures of dimension?}
\label{sec:counterexample}

In his monograph \cite{robinson11}, Robinson studies the problem of embedding a \emph{compact} subset $\Sigma$ of an infinite-dimensional (Hilbert or Banach) space into $\Rbb^m$. He considers different definitions of dimension, in particular, the Hausdorff dimension (denoted $dim_H$ hereafter) and the upper box-counting dimension. He shows that embedding the set $\Sigma$ into $\Rbb^m$ is possible as soon as ${\rm dim}_H(\Sigma-\Sigma) < m$ but that no modulus of continuity is possible for $L^{-1}$, \ie, the embedding is not necessarily stable. One should further assume that the upper box-counting dimension of $\Sigma$ is finite in order to find \emph{linear} embeddings with H\"older continuous inverses. 

In our work, we show that a stable embedding of a set $\Sigma$ into finite dimensions exists when the upper box-counting dimension on the \emph{normalised} secant set of $\Sigma$ is finite. Compared to \cite{robinson11} which uses the dimension of the secant set $\Sigma - \Sigma$, we are thus adding another constraint by normalising the secant set. An interesting question is thus whether we can change our measure of dimension and still guarantee the existence of stable linear embeddings or not. A first step in answering this question is checking if a finite upper box-counting dimension of $\SCal$ is a necessary condition to the existence of linear maps that satisfies the RIP. The following example shows that it is only a sufficient condition, which leaves room for a refinement of our results. This example is inspired by the construction of the ``orthogonal sequence'' in \cite{robinson11}.

Let $\HCal$ be the space of square-summable sequences, $\ell^2$, and define the correlated sequence $\Sigma = \{ \vec{x}_i \}_{i \geq 1}$ with $\vec{x}_i = r^i \, (\vec{e}_i + b \, \vec{e}_0)$, where $(\vec{e}_i)_{i \in \Nbb}$ is the standard orthonormal basis of $\HCal$, $r \in (0, 1)$ and $b > 0$. Now consider the linear map $L : \HCal \rightarrow \Rbb$ defined by $L(\vec{x}) = \scp{\vec{x}}{\vec{e}_0}$. Let us show that the linear map $L$ stably embeds the set $\Sigma$. Clearly,
\begin{align*}
\abs{\scp{\vec{x}_i - \vec{x}_j}{\vec{e}_0}} \leq \norm{\vec{x}_i - \vec{x}_j} \; \text{ for all } \vec{x}_i, \vec{x}_j \in \Sigma,
\end{align*}
so let us consider the lower isometry bound. We want to show the existence of an $\alpha>0$ such that
\begin{align*}
\abs{\scp{\vec{x}_i - \vec{x}_j}{\vec{e}_0}} \geq \alpha \, \norm{\vec{x}_i - \vec{x}_j}, \; \forall \vec{x}_i, \vec{x}_j \in \Sigma, j>i\geq 1,
\end{align*}
or equivalently,  
\begin{align*}
b^2 (r^i - r^j)^2 \geq \alpha^2 \; (r^{2i} +r^{2j} + b^2 (r^i - r^j)^2),\;  \forall j > i \geq 1.
\end{align*}
Re-arranging the terms gives
\begin{align*}
b^2 r^{2i} (1-r^{j-i})^2 \geq \alpha^2 \; r^{2i} (1+r^{2(j-i)} + b^2 ((1-r^{j-i})^2),
\end{align*}
for all $j > i \geq 1$. We can thus choose
\begin{align*}
\alpha^2 
& = \min_{j>i\geq1} \; \frac{b^2 \, (1-r^{j-i})^2}{1+r^{2(j-i)} + b^2 (1-r^{j-i})^2}\\
& > \frac{b^2 (1-r)^2}{(1+r^2 + b^2)} > 0.
\end{align*}
Hence, $L$ satisfies the RIP on $\SCal$, the normalised secant set of $\Sigma$:
\begin{align*}
\alpha \leq \abs{\scp{\vec{y}}{\vec{e}_0}} \leq 1,\; \text{ for all } \vec{y} \in \SCal.
\end{align*}
Consequently, $L$ stably embeds the set $\Sigma$. Let us now calculate the upper box-counting dimension of $\SCal$. Consider the following infinite sequence of vectors from $\SCal$,
\begin{align*}
\vec{v}_k 
:= \frac{\vec{x}_{2k} - \vec{x}_{2k+1}}{\norm{\vec{x}_{2k} - \vec{x}_{2k+1}}}
=\frac{\vec{e}_{2k} - r \vec{e}_{2k+1} + b (1-r)\vec{e}_0}{\sqrt{1+r^2+b^2(1-r)^2}},
\end{align*}
with $k \geq 1$. Note that each $\vec{v}_k$ is the only point in the sequence that has a non-zero component in the direction $\vec{e}_{2k}$. Therefore,
\begin{align*}
\norm{\vec{v}_{k} - \vec{v}_{k'}} \geq \frac{1}{\sqrt{1+r^2+b^2(1-r)^2}},\; \forall k \neq k'.
\end{align*}
Thus, for all $\epsilon < 2^{-1} [1+r^2+b^2(1-r)^2]^{-1/2}$, it is impossible to cover the set $\{\vec{v}_k\}_{k\geq1} \subset \SCal$ with a finite $\epsilon$-cover. Hence, $\dim(\SCal) = \infty$. We thus have a set $\SCal$ of infinite upper box-counting dimension for which there exists a linear map that satisfies the RIP. \emph{A finite upper box-counting dimension of the normalised secant set is thus not necessary for the existence of a stable linear embedding.}

Let us terminate this section by highlighting an important fact. We mentioned earlier that the upper box-counting dimension of $\SCal$ is often twice the upper box-counting of $\Sigma$ (or $\Sigma \cap \Sbb$ if $\Sigma$ is not bounded). It is however not always the case. Indeed, let us calculate the upper box-counting dimension of $\Sigma$. Let $\epsilon > 0$. We note that
\begin{align*}
\norm{\vec{x}_i} = r^i (1+b^2)^{1/2}.
\end{align*}
Let $i^*(\epsilon)$ be the smallest integer such that $r^{i^*(\epsilon)} (1+b^2)^{1/2} \leq \epsilon$. For $\epsilon$ small enough, we have $i^*(\epsilon)\geq2$. Hence, we can cover $\{x_i\}_{i \geq i^*(\epsilon)}$ with a single $\epsilon$-ball at $0$. Then, the remaining $i^*(\epsilon)-1$ points, $\{x_i\}_{i=1}^{i^*(\epsilon)-1}$ can be separately covered by at most $i^*(\epsilon)-1$ balls of radius $\epsilon$ centered at each $x_i$. This gives $N_{\Sigma}(\epsilon) \leq i^*(\epsilon)$. Then,
\begin{align*}
\frac{\log N_{\Sigma}(\epsilon)}{-\log \epsilon} \; \leq \; \frac{\log i^*(\epsilon)}{-(i^*(\epsilon)-1) \log r - \frac{1}{2} \log (1+b^2)}.
\end{align*}
Hence,
\begin{align*} 
\dim(\Sigma) 
& =  \limsup_{\epsilon  \rightarrow 0} \frac{\log N_{\Sigma}(\epsilon)}{-\log \epsilon} \\
& \leq \limsup_{i^{*} \rightarrow \infty} \frac{\log i^*}{-(i^*-1) \log r - \frac{1}{2} \log (1+b^2)} \\
& = 0.
\end{align*}
Since $\dim(\Sigma)$ cannot be negative, we have $\dim(\Sigma)  = 0$. We thus have an example of a set $\Sigma$ with a null upper box-counting dimension but whose normalised secant set has an infinite upper box-counting dimension. In general, one cannot directly deduce the dimension of $\SCal$ from the dimension of $\Sigma$.

\subsection{Related works with similar proof techniques}

A closely related work is the one of Dirksen \cite{dirksen14}. In this work, the ambient space is a (possibly infinite) Hilbert space and a generic theory for dimensionality reduction with subgaussian maps is presented. This work allows one to derive embedding results \emph{once a random linear map} $L$ is given. Theorem \ref{th:rip_2} can for example be derived using the generic theory developed in \cite{dirksen14}. In our work, we provide in addition a recipe for the construction of the linear map $L$ itself. Let us also note that one cannot directly recover the results presented in Section \ref{sec:rop} with the result presented in \cite{dirksen14}.

Another related work of Dirksen is \cite{dirksen14b}, that extends results presented in \cite{klartag05, mendelson07} \add{and where a technique to obtain tail bounds of the supremum of a stochastic process is presented. Let us highlight that these results can be another approach to obtain our generic theorem (Theorem \ref{th:main_theorem}). From a more technical point of view, \cite{dirksen14b} provides bounds for all $p$-th moment of the supremum of a stochastic process. Deviation inequalities immediately follows from these bounds. The proofs use the so-called generic chaining argument and the $\gamma$-functionals. Although the approximation of these functionals might not be easy, this method can produce sharper bounds than ours. In our proofs, we use more basic tools of probability theory with a another type of chaining argument to directly obtain the deviation inequalities.}

Related works also exist for embeddings of manifolds in finite ambient dimension \cite{baraniuk09, clarkson08, eftekhari14}. In particular, the result presented in Section \ref{sec:rip_hilbert} generalises the result presented in \cite{eftekhari14} to any model set with finite upper box-counting dimension. The proof technique used in \cite{eftekhari14} is also based on a chaining argument. In \cite{eftekhari14}, one of the main contribution is also the identification of the manifold properties that allow one to control the covering dimension. In \cite{mantzel14}, the same type of embedding result also appears but for signals that lie on a collection of continuously parameterised low-dimensional subspaces in $\Rbb^n$. 

\add{Even though the following result applies in a finite dimensional Euclidean space, let us mention that Oymak \etal\  proved in \cite{oymak15} that certain structured random matrices can embed any low-dimensional set of $\Rbb^n$. This is a particularly useful result for practical applications as it shows that matrices for which fast matrix-vector multiplication algorithms exist can be used to embed any set, as long as it has a small intrinsic dimension. It would be interesting in a future work to study how such computationally efficient matrices can be used in the construction of our linear map $L$.}

\section{Conclusion}
\label{sec:conclusion}

We presented a generic recipe to prove that a random linear map satisfies the RIP on arbitrary low dimensional sets $\SCal$ which lives in a Hilbert space. The proposed framework is general enough to take into account a large class of subsets $\SCal$ as well as structured and unstructured measurement processes. We also explained how to \emph{construct} random linear maps that satisfy the RIP with high probability in this general setting.

The linear map presented in Section \ref{sec:rip_hilbert} is built in two steps. The first step consists in a projection onto a well-chosen finite-dimensional subspace $\VCal_{\epsilon_*}$ and the second step uses a random subgaussian matrix to reduce the dimension. In order to obtain linear maps which are computationally more efficient, one can consider replacing the random matrix in the second step by a more structured one. One possibility is to use a variable density sampling technique, as we considered in \cite{puy15}. Note that the condition on $m$ presented in \cite{puy15} is not optimal. Indeed for $k$-sparse signals in $\Rbb^n$, the condition requires a number of measurements essentially proportional to $k^2$ to ensure that the RIP is satisfied. To understand the performance of the variable density sampling technique in practical applications, it will be important to determine under which additional structures this condition on $m$ can be improved.

We also saw that a finite upper box-counting dimension of the normalised secant set $\add{\mathcal{S}({\Sigma})}$ of $\Sigma$ is not necessary for the existence of a stable dimension reducing linear embedding of $\Sigma$. An interesting question is thus what appropriate measure of dimension of $\add{\mathcal{S}({\Sigma})}$ is both necessary and sufficient for the existence of a stable dimension reducing linear embedding of $\Sigma$.

\appendices

\section{Proof of Theorem \refMainTheorem}
\label{app:main_theorem}

The proof of Theorem \refMainTheorem\ is based on a chaining argument which is a powerful technique to obtain sharp bounds for the supremum of random processes \cite{talagrand96, clarkson08, rauhut10, eftekhari14}. Before proving Theorem \refMainTheorem, we need some preparations.

First, we notice that if Assumption \ref{as:dimension} holds then we can cover the set $\SCal$ with a finite number of balls. Furthermore, a bound on the number of balls sufficient to cover $\SCal$ is available for all radius $\epsilon \leq \epsilon_{\SCal}$. In the proof below, for each $j \geq 0$,
\begin{itemize}
\item $\CCal_j \subset \SCal$, denotes a minimal $(2^{-j} \epsilon_{\SCal})$-net for $\SCal$; 
\item $\pi_j$ denotes the mapping $\pi_j (\vec{x}) \in \argmin_{\vec{z} \in \CCal_j} \norm{\vec{x} - \vec{z}}$; 
\item $\DCal_j$ denotes the finite set $\{(\pi_{j+1}(\vec{x}), \pi_{j}(\vec{x})) \; \vert \; \vec{x} \in \SCal \}\subset \CCal_{j+1} \times \CCal_{j}$.
\end{itemize}
One can remark that the cardinality of $\CCal_j$ is $N_{\SCal}(2^{-j} \epsilon_{\SCal})$, that the cardinality of $\DCal_j$ is bounded above by $N_{\SCal}^2 (2^{-j-1} \epsilon_{\SCal})$, and that $\sup_{(\vec{y}, \vec{z}) \in \DCal_j} \norm{\vec{y} - \vec{z}}  \leq 2^{-j+1} \epsilon_{\SCal}$. 

Second, we remark that Assumption \ref{as:concentration} implies that for any fixed $\vec{x} \in \SCal \subset \Sbb$
\begin{align}
\label{eq:prob_bound_unique}
\Prob \left\{ \abs{h_{p}(\vec{x})} \geq \lambda \right\}
\leq \left\{
\begin{array}{l}
2 \ee^{- c_1 m \lambda^2}, \text{if}\ 0 \leq \lambda \leq \frac{c_2}{c_1},\\
2  \ee^{- c_1 m \lambda}, \text{if}\ \lambda \geq \frac{c_2}{c_1}.
\end{array}
\right.
\end{align}
Indeed, it suffices to take $\vec{y} = \vec{x}$ and $\vec{z} = \vec{0}$ in \refeq{eq:prob_bound_increment_1} and \refeq{eq:prob_bound_increment_2}. 

With these tools in hand, we can prove the following intermediate result from which Theorem \refMainTheorem\ follows.

\begin{lemma}
\label{lemma:chaining}
Let $(\vec{l}_1, \ldots, \vec{l}_m)$ be $m$ random vectors drawn from $\HCal^m$ according to a probability distribution $\mu$, $L~:~\HCal \rightarrow \Rbb^m$ be the linear map defined in \refeq{eq:generic_linear_map}, $\delta_p$ be the quantity defined in \refeq{eq:true_RIP_constant} and $\xi \in (0, 1)$. Define
\begin{gather*}
S_1(\epsilon_{\SCal}, \xi) :=  \sqrt{\log\left(\frac{2}{\xi} \cdot N_{\SCal}\left(\epsilon_{\SCal}\right)\right)},\\
S_2(\epsilon_{\SCal}, \xi) := \sum_{j \in \Nbb} 2^{-j+1} \sqrt{\log\left( \frac{2^{j+1}}{\xi} \cdot N_{\SCal}^2 \left( \frac{\epsilon_{\SCal}}{2^{j+1}} \right) \right)},\\
S_3(\epsilon_{\SCal}, \xi) := \sum_{j \in \Nbb} 2^{-j+1} \log\left( \frac{2^{j+1}}{\xi} \cdot N_{\SCal}^2 \left( \frac{\epsilon_{\SCal}}{2^{j+1}} \right) \right).
\end{gather*}
If Assumption \ref{as:concentration} holds, then
\begin{align*}
\delta_{p} \leq \frac{S_1(\epsilon_{\SCal}, \xi) + S_2(\epsilon_{\SCal}, \xi) \epsilon_{\SCal} }{\sqrt{c_1m}} + \frac{S_1^2(\epsilon_{\SCal}, \xi) + S_3(\epsilon_{\SCal}, \xi) \epsilon_{\SCal} }{c_2m}
\end{align*}
with probability at least $1 - 3\xi$.
\end{lemma}
\begin{IEEEproof}
We begin by establishing the telescopic sum expression:
\begin{align*}
h_{p}(\vec{x}) = h_{p}(\pi_{0}(\vec{x})) + \sum_{j=0}^\infty \left[h_{p}(\pi_{j+1}(\vec{x})) - h_{p}(\pi_{j}(\vec{x})) \right].
\end{align*}
The above equality holds because the linear maps of the form of $L$ are continuous with respect to $\norm{\cdot}$. Therefore, $\norm{L(\cdot)}_p^p$, $\Ebb \norm{L(\cdot)}_p^p$, and thus $h_p$ are also continuous with respect to $\norm{\cdot}$. Then,
\begin{align*}
& h_{p}(\vec{x}) - h_{p}(\pi_{0}(\vec{x})) - \sum_{j=0}^N \left[h_{p}(\pi_{j+1}(\vec{x})) - h_{p}(\pi_{j}(\vec{x})) \right]\\
& \hspace{30mm} = h_{p}(\vec{x}) - h_{p}(\pi_{N+1}(\vec{x})).
\end{align*}
As $\lim_{N \rightarrow \infty} \norm{\pi_{N+1}(\vec{x}) - \vec{x}} = 0$, we have $\lim_{N \rightarrow \infty} \abs{h_{p}(\vec{x}) - h_{p}(\pi_{N+1}(\vec{x}))} = 0$ by continuity. 

We continue with the triangle inequality which yields
\begin{align*}
\delta_p 
& = \sup_{\vec{x} \in \SCal } \abs{h_{p}(\vec{x})} \\
& \leq \sup_{\vec{x} \in \SCal } \abs{h_{p}(\pi_{0}(\vec{x}))} + \sum_{j=0}^\infty \; \sup_{\vec{x} \in \SCal } \abs{h_{p}(\pi_{j+1}(\vec{x})) - h_{p}(\pi_{j}(\vec{x}))} \\
& =  \max_{\vec{x}_0 \in \CCal_0} \abs{h_{p}(\vec{x}_0)} + \sum_{j=0}^{\infty} \max_{(\vec{y}, \vec{z}) \in \DCal_j} \abs{h_{p}(\vec{y}) - h_{p}(\vec{z})}.
\end{align*}
Let $a_j, b > 0$, with $j \in \Nbb$, be parameters whose values will be chosen later on.  The union bound yields
\begin{align}
\label{eq:prob_bound_chaining}
& \Prob \left\{ \delta_{p} \geq b + \sum_{j \geq 0} a_j \right\} \nonumber \\
& \hspace{2mm} \leq \Prob \left\{ \max_{\vec{x}_0 \in \CCal_0} \abs{h_{p}(\vec{x}_0)} \geq b  \right\} \nonumber \\
& \hspace{12mm} + \sum_{j \geq 0} \Prob \left\{  \max_{(\vec{y}, \vec{z}) \in \DCal_j} \abs{h_{p}(\vec{y}) - h_{p}(\vec{z})} \geq a_j  \right\} \nonumber \\
& \hspace{2mm} \leq N_{\SCal} \left( {\epsilon_{\SCal}} \right) \cdot \max_{\vec{x}_0 \in \CCal_0} \Prob \left\{  \abs{h_{p}(\vec{x}_0)} \geq b  \right\} \\
& \hspace{2mm}  + \sum_{j \geq 0} \Big[ N_{\SCal}^2 \left( \frac{\epsilon_{\SCal}}{2^{j+1}} \right) \cdot \max_{(\vec{y}, \vec{z}) \in \DCal_j} \Prob \left\{ \abs{h_{p}(\vec{y}) - h_{p}(\vec{z})} \geq a_j  \right\} \Big]. \nonumber
\end{align}
In the last step, we used the facts that the cardinality of $\CCal_0$ is $N_{\SCal} \left( {\epsilon_{\SCal}} \right)$, and that the cardinality of $\DCal_j$ is bounded above by $N_{\SCal}^2 \left( 2^{-j-1}\epsilon_{\SCal} \right)$.

We now bound the first term on the right-hand side (rhs) of \refeq{eq:prob_bound_chaining}. We recall that $\CCal_0 \subset \SCal$. Inequality \refeq{eq:prob_bound_unique} thus yields
\begin{align*}
\max_{\vec{x}_0 \in \CCal_0} \Prob \left\{  \abs{h_{p}(\vec{x}_0)} \geq b  \right\} 
\leq \left\{
\begin{array}{l}
2 \ee^{- c_1 m b^2}, \text{if}\ b \leq \frac{c_2}{c_1},\\
2  \ee^{- c_2 m b}, \text{if}\ b \geq \frac{c_2}{c_1}.
\end{array}
\right.
\end{align*}
If
\begin{align*}
\frac{S_1^2(\epsilon_{\SCal}, \xi)}{m} \leq \frac{c_2^2}{c_1},
\end{align*}
we choose
\begin{align*}
b := \frac{S_1(\epsilon_{\SCal}, \xi)}{\sqrt{c_1m}} = \sqrt{\frac{\log(2N_{\SCal} \left( {\epsilon_{\SCal}} \right)/\xi)}{c_1m} },
\end{align*}
and 
\begin{align*}
b := \frac{S_1^2(\epsilon_{\SCal}, \xi)}{c_2m} = \frac{\log(2N_{\SCal} \left( {\epsilon_{\SCal}} \right)/\xi)}{c_2m},
\end{align*}
otherwise. This choice yields
\begin{align*}
N_{\SCal} \left( {\epsilon_{\SCal}} \right) \cdot \max_{\vec{x}_0 \in \CCal_0} \Prob \left\{ \abs{h_{p}(\vec{x}_0)} \geq b \right\} \leq \xi,
\end{align*}
and
\begin{align*}
b \leq \frac{S_1(\epsilon_{\SCal}, \xi)}{\sqrt{c_1m}} + \frac{S_1^2(\epsilon_{\SCal}, \xi)}{c_2m},
\end{align*}
in both cases.

We continue by bounding the infinite sum on the rhs of \refeq{eq:prob_bound_chaining}. We notice that for $(\vec{y}, \vec{z}) \in \DCal_j$, we have $\norm{\vec{y} - \vec{z}} \leq 2^{-j+1} \epsilon_{\SCal}$. Assumption \ref{as:concentration} shows that
\begin{align}
\label{eq:bound_increment_chaining}
& \Prob \left\{\abs{h_{p}(\vec{y}) - h_{p}(\vec{z})} \geq a_j \right\} \nonumber \\
&  \hspace{6mm} = \Prob\left\{ \abs{h_{p}(\vec{y}) - h_{p}(\vec{z})} \geq \frac{ a_j 2^{j-1} }{\epsilon_{\SCal}} \cdot  \frac{\epsilon_{\SCal}}{2^{j-1}} \right\} \nonumber \\ 
&  \hspace{6mm} \leq \Prob\left\{ \abs{h_{p}(\vec{y}) - h_{p}(\vec{z})} \geq \frac{ a_j 2^{j-1}}{\epsilon_{\SCal}} \norm{\vec{y} - \vec{z}} \right\} \nonumber \\ 
&  \hspace{6mm} \leq \left\{
\begin{array}{l}
2 \; \ee^{- c_1 \, m \, (2^{j-1} a_j/\epsilon_{\SCal})^2}, \text{ if } \frac{a_j}{2^{-j+1} \epsilon_{\SCal}} \leq c_2/c_1, \\
2 \; \ee^{- c_2 \, m \, 2^{j-1} a_j/\epsilon_{\SCal}}, \text{ if } \frac{a_j}{2^{-j+1} \epsilon_{\SCal}} \geq c_2/c_1.
\end{array}
\right.
\end{align}
Define
\begin{align*}
\mathcal{J} := \left\{ j \in \Nbb \, \Big\vert \, \inv{m} \log\left(\frac{2^{j+1}}{\xi} \cdot N_{\SCal}^2 \left( \frac{\epsilon_{\SCal}}{2^{j+1}} \right) \right) \leq \frac{c_2^2}{c_1} \right\}.
\end{align*}
We choose
\begin{align*}
a_j := \frac{2^{-j+1} \epsilon_{\SCal}}{ \sqrt{c_1 m}} \sqrt{\log\left(\frac{2^{j+1}}{\xi} \cdot N_{\SCal}^2 \left( \frac{\epsilon_{\SCal}}{2^{j+1}} \right) \right)},
\end{align*}
for all $j \in \mathcal{J}$, and
\begin{align*}
a_j :=  \frac{2^{-j+1} \epsilon_{\SCal}}{c_2m} \log\left(\frac{2^{j+1}}{\xi} \cdot N_{\SCal}^2 \left( \frac{\epsilon_{\SCal}}{2^{j+1}} \right)\right),
\end{align*}
for all $j \in \Nbb \setminus \mathcal{J}$. Notice that, by definition of $\mathcal{J}$, we have
\begin{align*}
\frac{a_j}{2^{-j+1} \epsilon_{\SCal}} \leq \frac{c_2}{c_1}
\end{align*}
for all $j \in \mathcal{J}$, and
\begin{align*}
\frac{a_j}{2^{-j+1} \epsilon_{\SCal}} \geq \frac{c_2}{c_1},
\end{align*}
otherwise. Replacing $a_j$ by its expression in \refeq{eq:bound_increment_chaining} shows that
\begin{align*}
N_{\SCal}^2 \left( \frac{\epsilon_{\SCal}}{2^{j+1}} \right) \max_{(\vec{y}, \vec{z}) \in \DCal_j} \Prob \left\{ \abs{h_{p}(\vec{y}) - h_{p}(\vec{z})} \geq a_j  \right\} \leq 2^{-j} \xi
\end{align*}
for all $j \in \Nbb$. Therefore,
\begin{align*}
\sum_{j \geq 0} N_{\SCal}^2 \left( \frac{\epsilon_{\SCal}}{2^{j+1}} \right) \max_{(\vec{y}, \vec{z}) \in \DCal_j} \Prob \left\{ \abs{h_{p}(\vec{y}) - h_{p}(\vec{z})} \geq a_j  \right\} \leq 2 \xi.
\end{align*}
In total, the rhs of \refeq{eq:prob_bound_chaining} is thus bounded by $3\xi$. Finally, we have
\begin{align*}
\sum_{j \geq 0} a_j 
& = \frac{\epsilon_{\SCal}}{ \sqrt{c_1 m}} \sum_{j \in \mathcal{J}} 2^{-j+1} \sqrt{\log\left( \frac{2^{j+1}}{\xi} \cdot N_{\SCal}^2 \left( \frac{\epsilon_{\SCal}}{2^{j+1}} \right)\right)}\\
& + \frac{\epsilon_{\SCal}}{c_2m} \sum_{j \in \Nbb \setminus \mathcal{J}} 2^{-j+1} \log\left( \frac{2^{j+1}}{\xi} \cdot N_{\SCal}^2 \left( \frac{\epsilon_{\SCal}}{2^{j+1}} \right)\right)\\
& \leq \frac{\epsilon_{\SCal}}{ \sqrt{c_1 m}} \sum_{j \in \Nbb} 2^{-j+1} \sqrt{\log\left( \frac{2^{j+1}}{\xi} \cdot N_{\SCal}^2 \left( \frac{\epsilon_{\SCal}}{2^{j+1}} \right)\right)}\\
& + \frac{\epsilon_{\SCal}}{c_2m} \sum_{j \in \Nbb} 2^{-j+1} \log\left( \frac{2^{j+1}}{\xi} \cdot N_{\SCal}^2 \left( \frac{\epsilon_{\SCal}}{2^{j+1}} \right)\right)\\
& = \epsilon_{\SCal}\ \frac{S_2(\epsilon_{\SCal}, \xi)}{ \sqrt{c_1 m}} + \epsilon_{\SCal}\ \frac{S_3(\epsilon_{\SCal}, \xi)}{c_2m}.
\end{align*}

In summary, we have shown that
\begin{align*}
\delta_{p} 
& \leq b + \sum_{j \geq 0} a_j 
\end{align*}
with probability at least $1-3\xi$. Using the bounds on $b$ and $\sum_{j \geq 0} a_j$, we have
\begin{align*}
\delta_{p} \leq \frac{S_1(\epsilon_{\SCal}, \xi) + S_2(\epsilon_{\SCal}, \xi) \epsilon_{\SCal} }{\sqrt{c_1m}} + \frac{S_1^2(\epsilon_{\SCal}, \xi) + S_3(\epsilon_{\SCal}, \xi) \epsilon_{\SCal} }{c_2m}
\end{align*}
with at least the same probability.
\end{IEEEproof}

We can now prove Theorem \refMainTheorem. First, we need to compute the sums $S_1, S_2, S_3$ in the above lemma. It involves uninteresting computations which, for completeness, we detail in Appendix \ref{app:precomputations}. For $\xi \in (0, 1)$ and $\epsilon_{\SCal} \in (0, 1/2)$, if Assumption \ref{as:dimension} holds, these computations show that
\begin{align*}
S_1(\epsilon_{\SCal}, \xi) & \leq  \sqrt{\log\left({2}/{\xi}\right)} + \sqrt{s \log(1/\epsilon_{\SCal})}, \\
S_2(\epsilon_{\SCal}, \xi) & \leq  8 \sqrt{\log\left({2}/{\xi}\right)} + 8 \sqrt{2s \log(2)} + 4 \sqrt{2s \log(1/\epsilon_{\SCal})},\\
S_1^2(\epsilon_{\SCal}, \xi) & \leq  \log\left({2}/{\xi}\right) + s \log(1/\epsilon_{\SCal}), \\
S_3(\epsilon_{\SCal}, \xi) & \leq  8 \log\left({2}/{\xi}\right) + 16 s \log(2) + 8 s \log(1/\epsilon_{\SCal}).
\end{align*}
Therefore,
\begin{align*}
& \delta_{p}
\leq \sqrt{\frac{\log\left({2}/{\xi}\right)}{c_1m}} + \sqrt{\frac{s \log(1/\epsilon_{\SCal})}{c_1m}} \\
& + 8 \epsilon_{\SCal} \sqrt{\frac{\log\left({2}/{\xi}\right)}{c_1m}} + 8\epsilon_{\SCal} \sqrt{\frac{2s \log(2)}{c_1m}} + 4\epsilon_{\SCal} \sqrt{\frac{2s \log(1/\epsilon_{\SCal})}{c_1m}}\\
& + \frac{\log\left({2}/{\xi}\right)}{c_2m} + \frac{s \log(1/\epsilon_{\SCal})}{c_2m} \\
& + \frac{8 \epsilon_{\SCal}  \log\left({2}/{\xi}\right)}{c_2m} + \frac{16\epsilon_{\SCal} \; s \log(2)}{c_2m}  + \frac{8\epsilon_{\SCal} \; s \log(1/\epsilon_{\SCal})}{c_2m}.
\end{align*}
with probability at least $1-3\xi$.

Let $\delta \in (0, 1)$ and define $D = 1/\min(c_1, c_2)$. We remark that if
\begin{align}
\label{eq:cond_m_1}
m \geq \frac{16D}{\delta^2} \log(2/\xi),
\end{align}
then, since $\epsilon_{\SCal} < 1/2$,
\begin{gather*}
\sqrt{\frac{\log\left({2}/{\xi}\right)}{c_1m}} \leq \delta,\
8 \epsilon_{\SCal} \sqrt{\frac{\log\left({2}/{\xi}\right)}{c_1m}} \leq \delta,\\
\frac{\log\left({2}/{\xi}\right)}{c_2m} \leq \delta^2 \leq \delta, \
\text{and }\ 8 \epsilon_{\SCal} \; \frac{\log\left({2}/{\xi}\right)}{c_2m} \leq \delta^2 \leq \delta.
\end{gather*}
Similarly, if
\begin{align}
\label{eq:cond_m_2}
m \geq \frac{8D}{\delta^2} \; s \; \log(1/\epsilon_{\SCal}),
\end{align}
then, since $\epsilon_{\SCal} < 1/2$,
\begin{gather*}
\sqrt{\frac{s \log(1/\epsilon_{\SCal})}{c_1m}} \leq \delta,\
4\epsilon_{\SCal} \sqrt{\frac{2s \log(1/\epsilon_{\SCal})}{c_1m}} \leq \delta,\\
\frac{s \log(1/\epsilon_{\SCal})}{c_2m} \leq \delta^2 \leq \delta,\
\text{and }\ \frac{8\epsilon_{\SCal}s \log(1/\epsilon_{\SCal})}{c_2m} \leq \delta^2 \leq \delta.
\end{gather*}
Finally, if
\begin{align}
\label{eq:cond_m_3}
m \geq \frac{32D}{\delta^2} \; s \; \log(2),
\end{align}
then, since $\epsilon_{\SCal} < 1/2$,
\begin{gather*}
8\epsilon_{\SCal} \sqrt{\frac{2s \log(2)}{c_1m}} \leq \delta\
\text{ and }\ \frac{16\epsilon_{\SCal} s \log(2)}{c_2m} \leq \delta^2 \leq \delta.
\end{gather*}

To terminate the proof, we notice that if
\begin{align*}
m \geq \frac{D}{\delta^2} \max\left\{s \log\left(\inv{\epsilon_{\SCal}}\right),\  \log\left(\frac{2}{\xi}\right) \right\},
\end{align*}
with $D:= 32/\min(c_1, c_2)$, holds then \refeq{eq:cond_m_1}, \refeq{eq:cond_m_2}, \refeq{eq:cond_m_3} all hold. Under this condition, we have
$\delta_{p} \leq 10\delta$, with probability at least $1-3\xi$. Two change of variables ($\xi' = 3\, \xi$, $\delta' = 10\, \delta$) prove the theorem with $C=3200$.

\section{Evaluation of $S_1$, $S_2$, $S_3$ of Lemma \ref{lemma:chaining}}
\label{app:precomputations}

To estimates the sums $S_1$, $S_2$, $S_3$, we start by noticing that $\sum_{j \geq 0} j 2^{-j} = 2$ (see, \eg, Section 4.2.3 in \cite{riley06}). We precompute, for $\xi \in (0, 1)$, 
\begin{align*}
& \sum_{j = 0}^{\infty} 2^{-j} \sqrt{\log\left({2^{j+1}}/{\xi}\right)} \leq \sum_{j = 0}^{\infty} 2^{-j} \sqrt{j \log\left(2\right) + \log\left({2}/{\xi}\right)} \\
& \hspace{15mm} \leq \sum_{j = 0}^{\infty} 2^{-j} \sqrt{j \log\left(2\right)} + \sum_{j = 0}^{\infty} 2^{-j} \sqrt{\log\left({2}/{\xi}\right)} \\
& \hspace{15mm} \leq \sqrt{\log\left(2\right)} \sum_{j = 0}^{\infty} 2^{-j} j + 2 \sqrt{\log\left({2}/{\xi}\right)} \\
& \hspace{15mm} = 2 \left[ \sqrt{\log\left(2\right)} + \sqrt{\log\left({2}/{\xi}\right)} \right] \leq 4 \sqrt{\log\left({2}/{\xi}\right)}.
\end{align*}
Similarly,
\begin{align*}
\sum_{j = 0}^{\infty} 2^{-j} \log\left({2^{j+1}}/{\xi}\right) \leq 2  \left[ \log(2) + \log \left({2}/{\xi} \right) \right] \leq 4 \log \left({2}/{\xi} \right).
\end{align*}
We also have, for $\epsilon_{\SCal} \in (0, 1/2)$,
\begin{align*}
& \sum_{j = 0}^{\infty} 2^{-j} \sqrt{\log\left( 2^{2(j+1)s} \epsilon_{\SCal}^{-2s} \right)}\\
& \hspace{8mm} = \sum_{j = 0}^{\infty} 2^{-j} \sqrt{2(j+1)s \log\left( 2 ) + 2s \log(1/\epsilon_{\SCal}) \right)}\\
& \hspace{8mm} \leq \sqrt{2s \log(2)} \sum_{j = 0}^{\infty} 2^{-j} \sqrt{j+1} + 2 \sqrt{2s \log(1/\epsilon_{\SCal})}\\
& \hspace{8mm} \leq \sqrt{2s \log(2)} \sum_{j = 0}^{\infty} 2^{-j} (j+1) + 2 \sqrt{2s \log(1/\epsilon_{\SCal})}\\
& \hspace{8mm} = 4\sqrt{2s \log(2)} + 2 \sqrt{2s \log(1/\epsilon_{\SCal})}.
\end{align*}
Similarly,
\begin{align*}
\sum_{j = 0}^{\infty} 2^{-j} \log\left( 2^{2(j+1)s} \epsilon_{\SCal}^{-2s} \right) \leq 8s \log(2) + 4s \log(1/\epsilon_{\SCal}).
\end{align*}

If Assumption \ref{as:dimension} holds, we have
\begin{align*}
N_{\SCal} \left( \frac{\epsilon_{\SCal}}{2^{j+1}} \right) \leq 2^{(j+1)s} \epsilon_{\SCal}^{-s},
\end{align*}
for all $j \in \Nbb$. Therefore, using the pre-computations above, we obtain
\begin{align*}
S_2(\epsilon_{\SCal}, \xi) & \leq  8 \sqrt{\log\left({2}/{\xi}\right)} + 8 \sqrt{2s \log(2)} + 4 \sqrt{2s \log(1/\epsilon_{\SCal})},
\end{align*}
and
\begin{align*}
S_3(\epsilon_{\SCal}, \xi) & \leq  8 \log\left({2}/{\xi}\right) + 16 s \log(2) + 8s \log(1/\epsilon_{\SCal}).
\end{align*}
We also have
\begin{align*}
S_1^2(\epsilon_{\SCal}, \xi) & \leq  \log\left({2}/{\xi}\right) + s \log(1/\epsilon_{\SCal}),
\end{align*}
and
\begin{align*}
S_1(\epsilon_{\SCal}, \xi) & \leq  \sqrt{\log\left({2}/{\xi}\right)} + \sqrt{s \log(1/\epsilon_{\SCal})}.
\end{align*}
%

\section{Proof of Theorem \ref{th:rip_1} and Theorem \ref{th:rip_2}}
\label{app:proof_theorem_rip_1_2}

\subsection{Basic tools}
In this section, we use several properties of subgaussian and subexponential random vectors/variables (see Definition \ref{def:subexp} and Definition \ref{def:subgauss}). We let the reader refer to, \eg, \cite{vershynin12} for more information about them. We recall here one useful property.
\begin{itemize} 
\item If $X$ is a subexponential random variable then so is $X~-~\Ebb X$, and we have $\norm{X - \Ebb X}_{\Psi_1} \leq 2 \norm{X}_{\Psi_1}$ (Remark $5.18$, \cite{vershynin12}).
\end{itemize}

We will also use the following Bernstein-type inequality for subexponential random variables.
\begin{lemma}[Corollary $5.17$, \cite{vershynin12}]
\label{lemma:bernstein}
There exists an absolute constant $c>0$ such that for independent centered subexponential random variables $X_1, \ldots, X_m$ with subexponential norm bounded by $K > 0$, we have, for every $0 \leq t \leq K$,
\begin{gather*}
\Prob\left\{ \inv{m} \abs{\sum_{i=1}^m X_i} \geq t \right\} \leq 2\ \ee^{- c m t^2/K^2},
\end{gather*}
and, for every $t \geq K$,
\begin{gather*}
\Prob\left\{ \inv{m} \abs{\sum_{i=1}^m X_i} \geq t \right\} \leq 2\ \ee^{- c m t/K}.
\end{gather*}
\end{lemma}
%

\subsection{Proof of Theorem \ref{th:rip_1}}

The independent random vectors $a_1, \ldots, a_m \in \Rbb^d$ are assumed to be such that
\begin{align*}
\Lambda_1 & := \sup_{\substack{\vec{x} \in \{ \SCal - \SCal\} \cup  \SCal \\ \vec{x} \neq \vec{0} }} \left\{ {\norm{a_i^\adjoint b(\vec{x})}_{\Psi_1}}/{\norm{b(\vec{x})}_{b}} \right\} < +\infty. 
\end{align*}
Therefore, for all $p \geq 1$, we have
\begin{align}
\label{eq:bound_p_moment_subexp}
\Ebb \left(\abs{a_i^\adjoint b(\vec{x})}^p \right)^{1/p} \; \leq \; \Lambda_1 \; p \; \norm{b(\vec{x})}_{b},
\end{align}
for any $\vec{x}$ in $\{ \SCal - \SCal\} \cup \SCal$.

Let $\vec{y}, \vec{z}$ be two fixed vectors in $\SCal \cup \{ \vec{0} \}$. Define $X := \sum_{i=1}^m X_i$ where
\begin{align*}
X_i := \abs{a_i^\adjoint b(\vec{y})} - \abs{a_i^\adjoint b(\vec{z})} - \Ebb(\abs{a_i^\adjoint b(\vec{y})} - \abs{a_i^\adjoint b(\vec{z})}).
\end{align*}
It is clear that $X$ is a sum of $m$ independent centered random variables. To use Lemma~\ref{lemma:bernstein}, we need to show that these variables are subexponential and bound their subexponential norm. We have
\begin{align*}
\left(\Ebb \big\vert{\abs{a_i^\adjoint b(\vec{y})} - \abs{a_i^\adjoint b(\vec{z})}}\big\vert^p \right)^{1/p} 
& \leq \left( \Ebb \; \abs{a_i^\adjoint (b(\vec{y}) - b(\vec{z}))}^p \right)^{1/p} \\
& = \left( \Ebb \; \abs{a_i^\adjoint (b(\vec{y} - \vec{z}))}^p \right)^{1/p}\\
& \leq \Lambda_1 \; p \; \norm{b(\vec{y} - \vec{z})}_{b}\\
& \leq \Lambda_1 \; p \; \norm{\vec{y} - \vec{z}}.
\end{align*}
The second step follows from the linearity of $b$. The third step follows from \refeq{eq:bound_p_moment_subexp} and the fact that $\vec{y} - \vec{z} \in \{ \SCal - \SCal\} \cup \SCal$. The last step follows from inequality \refeq{eq:upper_bound_proj}. 

We deduce that $X_1, \ldots, X_m$ are independent centered subexponential random variables with subexponential norm bounded by $2 \Lambda_1 \norm{\vec{y} - \vec{z}}$. Observe that
\begin{align*}
h_{1}(\vec{y}) - h_{1}(\vec{z}) = \inv{m} \; \sum_{i=1}^m X_i.
\end{align*}
Lemma~\ref{lemma:bernstein} with $K = 2 \Lambda_1 \norm{\vec{y} - \vec{z}}$ yields 
\begin{align*}
\Prob \left\{ \abs{h_{1}(\vec{y}) - h_{1}(\vec{z})} \geq \lambda \norm{\vec{y} - \vec{z}} \right\}
\leq 2 \ee^{- c_1 m \lambda^2},
\end{align*}
for every $0 \leq \lambda \leq 2 \Lambda_1$, and
\begin{align*}
\Prob \left\{ \abs{h_{1}(\vec{y}) - h_{1}(\vec{z})} \geq \lambda \norm{\vec{y} - \vec{z}} \right\}
\leq 2  \ee^{- c_2 m \lambda},
\end{align*}
for every $\lambda \geq 2 \Lambda_1$, with $c_1 = c/(4 \Lambda_1^2)$ and $c_2 = c/(2\Lambda_1)$. Note that $c_2/c_1 = 2 \Lambda_1$. Hence, Assumption \ref{as:concentration} is satisfied. Theorem \ref{th:main_theorem} terminates the proof.

\subsection{Proof of Theorem \ref{th:rip_2}}

The independent random vectors $a_1, \ldots, a_m \in \Rbb^d$ are assumed to be such that
\begin{align*}
\Lambda_2 & := \sup_{\substack{\vec{x} \in \{ \SCal - \SCal\} \cup  \SCal \\ \vec{x} \neq \vec{0} }} \left\{ {\norm{a_i^\adjoint b(\vec{x})}_{\Psi_2}}/{\norm{b(\vec{x})}_{b}} \right\} < +\infty
\end{align*}
Therefore, for all $p \geq 1$, we have
\begin{align}
\label{eq:bound_p_moment_subgauss}
\Ebb \left(\abs{a_i^\adjoint b(\vec{x})}^p \right)^{1/p} \; \leq \; \Lambda_2 \; \sqrt{p} \; \norm{b(\vec{x})}_{b},
\end{align}
for any $\vec{x}$ in $\{ \SCal - \SCal\} \cup \SCal$.

Let $\vec{y}, \vec{z}$ be two fixed vectors in $\SCal \cup \{ \vec{0} \}$. Define $X := \sum_{i=1}^m X_i$ where
\begin{align*}
X_i := \abs{a_i^\adjoint b(\vec{y})}^2 - \abs{a_i^\adjoint b(\vec{z})}^2 - \Ebb(\abs{a_i^\adjoint b(\vec{y})}^2 - \abs{a_i^\adjoint b(\vec{z})}^2).
\end{align*}
It is clear that $X$ is a sum of $m$ independent centered random variables. To use Lemma~\ref{lemma:bernstein}, we need to show that these variables are subexponential and bound their subexponential norm. We have
\begin{align*}
& \left(\Ebb \abs{\abs{a_i^\adjoint b(\vec{y})}^2 - \abs{a_i^\adjoint b(\vec{z})}^2}^p \right)^{1/p} \\
& \hspace{2mm} = \left(\Ebb \left(\abs{\vec a_i^\adjoint (b(\vec{y}) + b(\vec{z}))}^p \abs{\vec a_i^\adjoint (b(\vec{y}) - b(\vec{z}))}^p\right) \right)^{1/p} \\
& \hspace{2mm} \leq \left(\Ebb\abs{\vec a_i^\adjoint b(\vec{y} + \vec{z})}^{2p} \right)^{\inv{2p}} \left( \Ebb \abs{\vec a_i^\adjoint b(\vec{y} - \vec{z})}^{2p} \right)^{\inv{2p}}.
\end{align*}
In the last step, we used the Cauchy-Schwarz inequality and the linearity of $b$. Using \refeq{eq:bound_p_moment_subgauss} and the facts that\footnote{Remark that $\SCal$ is symmetric. Therefore $\SCal + \SCal = \SCal - \SCal$.} $(\vec{y} \pm \vec{z}) \in \{ \SCal - \SCal\} \cup \SCal$, we obtain
\begin{align*}
\left(\Ebb\abs{\vec a_i^\adjoint b(\vec{y} \pm \vec{z})}^{2p} \right)^{\inv{2p}}
& \leq \Lambda_2 \; \sqrt{2p} \; \norm{b(\vec{y} \pm \vec{z})}_b \\
& \leq \Lambda_2 \; \sqrt{2p} \; \norm{\vec{y} \pm \vec{z}} \\
& \leq 
\left\{
\begin{array}{ll}
2\sqrt{2} \; \Lambda_2 \; \sqrt{p}, & \text{for } \vec{y} + \vec{z},\\
\sqrt{2} \; \Lambda_2 \; \sqrt{p} \; \norm{\vec{y} - \vec{z}}, & \text{for } \vec{y} - \vec{z}.\\
\end{array}
\right.
\end{align*}
In the second step, we used inequality \refeq{eq:upper_bound_proj}. In the last step, we used the fact that $\norm{\vec{y}} \leq 1$ and $\norm{\vec{z}} \leq 1$. We thus have
\begin{align*}
\left(\Ebb \abs{\abs{a_i^\adjoint \vec{y}}^2 - \abs{a_i^\adjoint \vec{z}}^2}^p \right)^{1/p}  \leq  4 \Lambda_2^2 \; p \; \norm{\vec{y} - \vec{z}}.
\end{align*}
Therefore, $X_1, \ldots, X_m$ are independent centered subexponential random variables with subexponential norm bounded by $8\Lambda_2^2\norm{\vec{y} - \vec{z}}$. Observe that
\begin{align*}
h_{2}(\vec{y}) - h_{2}(\vec{z}) = \inv{m} \; \sum_{i=1}^m X_i.
\end{align*}
Lemma~\ref{lemma:bernstein} with $K = 8\Lambda_2^2\norm{\vec{y} - \vec{z}}$ yields 
\begin{align*}
\Prob \left\{ \abs{h_{2}(\vec{y}) - h_{2}(\vec{z})} \geq \lambda \norm{\vec{y} - \vec{z}} \right\}
\leq 2 \ee^{- c_1 m \lambda^2},
\end{align*}
for every $0 \leq \lambda \leq 8\Lambda_2^2$, and
\begin{align*}
\Prob \left\{ \abs{h_{2}(\vec{y}) - h_{2}(\vec{z})} \geq \lambda \norm{\vec{y} - \vec{z}} \right\}
\leq 2  \ee^{- c_2 m \lambda},
\end{align*}
for every $\lambda \geq 8\Lambda_2^2$ with $c_1 = c/(64 \Lambda_2^4)$ and $c_2 = c/(8\Lambda_2^2)$. Note that $8\Lambda_2^2 = c_2/c_1$. Hence, Assumption \ref{as:concentration} is satisfied. Theorem \ref{th:main_theorem} terminates the proof.

%
\section{Random rank-one projections}

\subsection{Estimation of $\Lambda_1$ from Section \ref{sec:rop}}

Let $\ma{M}$ be a fixed matrix in $\Rbb^{n_1 \times n_2}$ and $a \in \Rbb^{n_1}$, $b \in \Rbb^{n_2}$ be vectors whose entries are independent copies of the random variable $N$ or $P$, defined in \refeq{eq:sparse_random_variable}. In \cite{cai15}, the authors show that $a^\adjoint \ma{M} b$ is a subexponential random variable. In this section, we estimate the parameter $\Lambda_1$ needed in Recipe \ref{alg:recipe_rip_bis}. This estimation follows from the lemma below, whose proof starts from intermediate results presented in \cite{cai15}. 

Before presenting the lemma, we recall the definition of the double factorial for odd numbers: $(2k-1)!! = (2k)!/(2^k \, k!)$ for all $k\geq1$. Using Stirling's formula for the factorial, we obtain
\begin{align}
\label{eq:stirling_double_factorial}
(2k-1)!! 
= \frac{\sqrt{2\pi2k} \,(2k/\ee)^{2k} \,\ee^{\lambda_{2k}}}{2^k\sqrt{2\pi k} \, (k/\ee)^{k} \, \ee^{\lambda_{k}}}  = \sqrt{2} \; 2^{k} (k/\ee)^{k} \; \ee^{\lambda_{2k}-\lambda_{k}},
\end{align}
where $1/(12k+1) \leq \lambda_{k} \leq 1/(12k)$.

\begin{lemma}
\label{lemma:subexp_ROP}
Let $a \in \Rbb^{n_1}$ and $b \in \Rbb^{n_2}$ be vectors whose entries are independent copies of a random variable $X \in \Rbb$. Define
\begin{align*}
\alpha_X := \sup_{k \geq 1} \left[ \frac{\Ebb \,  X^{2k}}{(2k-1)!!} \right]^{1/(2k)}.
\end{align*}
Assume that $1 \leq \alpha_X < \infty$. For any fixed matrix $\ma{M}$ in $\Rbb^{n_1 \times n_2}$ and all $p\geq1$,
\begin{align*}
\left(\Ebb \abs{a^\adjoint \ma{M} b}^{p}\right)^{1/p} 
& \leq 2^{3/(2p)} \; \ee^{-1} \; p \; \alpha_X^2 \; \norm{\ma{M}}_{\rm Frob}.
\end{align*}
As a consequence, $a^\adjoint \ma{M} b$ is a subexponential random variable with subexponential norm bounded by $C \, \alpha_X^2 \norm{\ma{M}}_{\rm Frob}$, where $C = 2^{3/2} \; \ee^{-1} \approx 1.04>1$.
\end{lemma}

The proof of this lemma is given below. From the above result, we see that we only need to estimate $\alpha_X$ with $X = N$ and $X = P$ to estimate the constant $\Lambda_1$ in Section \ref{sec:rop}.

\begin{itemize}
\item \textbf{[Case where $X = N$]} As $N$ is a standard random variable, we have $\Ebb \,  N^{2k} = (2k-1)!!$. Therefore, $\alpha_N = 1$ and $\Lambda_1 \leq C$ where $C>1$ is an absolute constant.
\item \textbf{[Case where $X = P$]} It is easy to notice that $\Ebb \, P^{2k} = q^k/q$. Using  \refeq{eq:stirling_double_factorial} and the fact that $\ee^{\lambda_{2k}-\lambda_{k}} \geq 1/2$, we obtain
\begin{align*}
\left[ \frac{\Ebb P^{2k}}{(2k-1)!!} \right]^{1/(2k)}
& \leq \left[ \sqrt{2} \; 2^{-k} \; (k/\ee)^{-k} \;\; q^k/q\right]^{1/(2k)}\\
& \leq \left[ \sqrt{2} \; q^{-1} \; 2^{-k} \; (k/\ee)^{-k} \right]^{1/(2k)}\ q^{1/2}.
\end{align*}
Therefore, $\alpha_P \leq 1.39 \, \sqrt{q}$ and $\Lambda_1 \leq C \, \sqrt{q}$ where $C>1$ is an absolute constant.
\end{itemize}
\begin{IEEEproof}
We start with inequality ($0.20$) in the supplementary material of \cite{cai15}, which shows that, for all $k \geq 1$,
\begin{align*}
\Ebb \, \abs{a^\adjoint \ma{M} b}^{2k} \leq \alpha_X^{4k} \left[ (2k-1)!! \right]^2 \norm{\ma{M}}_{\rm Frob}^{2k}.
\end{align*}
Using \refeq{eq:stirling_double_factorial} and the fact that $\ee^{\lambda_{2k}-\lambda_{k}}~\leq~1$, we obtain $(2k-1)!! \leq \sqrt{2} \; 2^{k} (k/\ee)^{k}$. Therefore,
\begin{align*}
\Ebb \, \abs{a^\adjoint \ma{M} b}^{2k} \leq 2 \; (2\alpha_X^{2}/\ee)^{2k} \; k^{2k} \norm{\ma{M}}_{\rm Frob}^{2k},
\end{align*}
for all $k~\geq~1$. To shorten notations, we define 
\begin{align*}
C_{\alpha_X, \ma{M}}~:=~(2\alpha_X^{2}/\ee) \; \norm{\ma{M}}_{\rm Frob}.
\end{align*}
We thus have
\begin{align*}
\Ebb \, \abs{a^\adjoint \ma{M} b}^{2k} \leq 2 \, C_{\alpha_X, \ma{M}}^{2k} \; k^{2k}.
\end{align*}

We now follow exactly the interpolation technique used in the proof of Corollary $8.7$ in \cite{foucart13}. Let $\theta \in [0, 1]$. Then,
\begin{align*}
& \Ebb \abs{a^\adjoint \ma{M} b}^{2(k + \theta)} \\
& \hspace{5mm} \leq \left(\Ebb \abs{a^\adjoint \ma{M} b}^{2k}\right)^{1 - \theta} \, \left(\Ebb \abs{a^\adjoint \ma{M} b}^{2k + 2}\right)^{\theta} \\
& \hspace{5mm} \leq \left(2 \, C_{\alpha_X, \ma{M}}^{2k} \; k^{2k} \right)^{1 - \theta} \; \left(2 \, C_{\alpha_X, \ma{M}}^{2(k + 1)} \; (k+1)^{2(k +1)}\right)^{\theta} \\
& \hspace{5mm} = 2 \, C_{\alpha_X, \ma{M}}^{2(k+\theta)} \; k^{2k(1 - \theta)} \; (k+1)^{2(k+1)\theta} \\
& \hspace{5mm} = 2 \, C_{\alpha_X, \ma{M}}^{2(k+\theta)} \; \left[k^{1-\theta} \, (k+1)^{\theta} \right]^{2(k+\theta)} \left( \frac{k+1}{k} \right)^{2\theta(1-\theta)} \\
& \hspace{5mm} \leq 2 \, C_{\alpha_X, \ma{M}}^{2(k+\theta)} \; \left[k + \theta \right]^{2(k+\theta)} \left( \frac{k+1}{k} \right)^{2\theta(1-\theta)} \\
& \hspace{5mm} \leq 2\sqrt{2} \; C_{\alpha_X, \ma{M}}^{2(k+\theta)} \; \left[k + \theta \right]^{2(k+\theta)}
\end{align*}
H\"older's inequality yielded the first step. In the second step from below, we used the fact that
\begin{align*}
(1-\theta) \log(k) + \theta \log(k+1) \leq \log\left((1-\theta) k + \theta (k+1)\right),
\end{align*}
which follows from the concavity of the logarithm. In the last step, we used the facts that $(k+1)/k \leq 2$ and that $2\theta(1-\theta) \leq 1/2$. Taking $p = 2(k+\theta)$ proves the lemma for $p \geq 2$.

For $0 < p \leq 2$, we have
\begin{align*}
(\Ebb \abs{a^\adjoint \ma{M} b}^{p})^{1/p} 
& \leq (\Ebb \abs{a^\adjoint \ma{M} b}^{2})^{1/2} = \norm{\ma{M}}_{\rm Frob}\\
& \leq 2^{3/(2p)} \; \ee^{-1} \; p \; \alpha_X^2 \norm{\ma{M}}_{\rm Frob}.
\end{align*}
In the first step, we used the fact that for any random variable $X$, $(\Ebb \abs{X}^p)^{1/p} \leq (\Ebb \abs{X}^q)^{1/q}$ for $0<p \leq q$. The last inequality follows from the fact that the minimum of $f(p) = 2^{3/(2p)} \ee^{-1} \; p$ is attained at $p_0 = 3\log(2)/2$ and that $\alpha_X^{2} f(p_0) \approx 1.0397 \; \alpha_X^{2} \geq 1$. The lemma is thus proved for all $p > 0$.
\end{IEEEproof}
%

\subsection{Bound on $\underline{\delta}_1$}

Let us first introduce the following lemma which bounds \Ebb \abs{a^\adjoint x} for isotropic subexponential random vectors $a \in\Rbb^{n}$ and any fixed vector $x \in \Rbb^{n}$.

\begin{lemma} 
\label{lemma:subexp_bound_expected}
Let $a \in\Rbb^{n}$ be a random vector such that for any fixed vector $x \in \Rbb^{n}$,
\begin{align*}
\Ebb \abs{a^\adjoint x}^2 = \norm{x}_{2}^2\ \text{ and } \norm{a^\adjoint x}_{\Psi_1} \leq C \norm{x}_{2},
\end{align*}
where $C\geq2$ is a constant. Then,
\begin{align*}
\frac{\norm{x}_{2}}{2\ee^3 \, C \, \left(1+\log(C) \right)} \leq \Ebb \abs{a^\adjoint x} \leq \norm{x}_{2}.
\end{align*}
\end{lemma}

The proof of this lemma is given below. To estimate $\underline{\delta}_1$, we thus notice that we only need to estimate the constant $C$ in the above lemma. For rank-one projections, Lemma \ref{lemma:subexp_ROP} indicates that $C \leq O(\alpha_X^2)$.

\begin{IEEEproof}
The upper bound follows from Jensen's inequality,
\begin{align*}
\Ebb \abs{a^\adjoint x} \leq \sqrt{\Ebb \abs{a^\adjoint x}^2} = \norm{x}_{2}.
\end{align*}

For the lower bound, we start with H\"older's inequality. Let $\theta \in (0, 1)$, we obtain
\begin{align*}
\Ebb \abs{a^\adjoint x}^2
& = \Ebb \left(\abs{a^\adjoint x}^{1-\theta} \abs{a^\adjoint x}^{1+\theta} \right)\\
& \leq \left(\Ebb \abs{a^\adjoint x}\right)^{1-\theta} \left(\Ebb \abs{a^\adjoint x}^{(1+\theta)/\theta} \right)^{\theta}.
\end{align*}
Therefore,
\begin{align*}
\Ebb \abs{a^\adjoint x}
\geq \frac{\left(\Ebb \abs{a^\adjoint x}^2\right)^{1/(1-\theta)}}{\left(\Ebb \abs{a^\adjoint x}^{(1+\theta)/\theta}\right)^{\theta/(1-\theta)}}
\end{align*}
Using the fact that $a^\adjoint x$ has a subexponential norm bounded by $C \norm{x}_{2}$, we obtain
\begin{align*}
\Ebb \abs{a^\adjoint x}^{(1+\theta)/\theta} \leq \left(\frac{1+\theta}{\theta}\right)^{(1+\theta)/\theta} C^{(1+\theta)/\theta} \norm{x}_{2}^{(1+\theta)/\theta}.
\end{align*}
Therefore,
\begin{align*}
\Ebb \abs{a^\adjoint x}
& \geq \frac{\norm{x}_{2}^{2/(1-\theta)}}{\left[(1+\theta)/\theta \cdot C \cdot \norm{x}_{2} \right]^{(1+\theta)/(1-\theta)}} \\
& \geq \frac{\norm{x}_{2}}{\left[C \cdot (1+\theta)/\theta \right]^{(1+\theta)/(1-\theta)}}.
\end{align*}
The lower bound is valid for any $\theta \in (0, 1)$. Therefore,
\begin{align*}
\Ebb \abs{a^\adjoint x}
\geq \frac{\norm{x}_{2}}{\beta},
\end{align*}
where 
\begin{align*}
\beta := \inf_{\theta \in (0, 1)} \left[C \cdot (1+\theta)/\theta \right]^{(1+\theta)/(1-\theta)}.
\end{align*}
Let $g$ be the function defined by
\begin{align*}
g(\theta) =  \frac{1+\theta}{1-\theta} \log(C) +   \frac{1+\theta}{1-\theta} \log\left(1+\inv{\theta}\right).
\end{align*}
We have $\log(\beta) = \inf_{\theta \in (0, 1)} g(\theta)$. One can check that the minimum of $g$ is attained at $\theta^*$ satisfying
\begin{align*}
\theta_1 := \inv{5+3\log(C)} \; \leq \; \theta^* \; \leq \; \theta_2 := \inv{1+2\log(C)}.
\end{align*}
This gives
\begin{align*}
g(\theta_2) = \left[1 + \inv{\log(C)} \right] \left[ \log(C) +  \log\left(2+2\log(C)\right) \right] \geq g(\theta^*).
\end{align*}
For $C\geq2$, we have
\begin{align*}
3 \geq \frac{\log(C) +  \log\left(2+2\log(C)\right)}{\log(C)}.
\end{align*}
Therefore,
\begin{align*}
2\ee^3 \, C \, \left(1+\log(C) \right) \geq \beta.
\end{align*}
\end{IEEEproof}
%

\bibliographystyle{IEEEtran}
\bibliography{biblio}

\begin{thebibliography}{10}
\providecommand{\url}[1]{#1}
\csname url@samestyle\endcsname
\providecommand{\newblock}{\relax}
\providecommand{\bibinfo}[2]{#2}
\providecommand{\BIBentrySTDinterwordspacing}{\spaceskip=0pt\relax}
\providecommand{\BIBentryALTinterwordstretchfactor}{4}
\providecommand{\BIBentryALTinterwordspacing}{\spaceskip=\fontdimen2\font plus
\BIBentryALTinterwordstretchfactor\fontdimen3\font minus
  \fontdimen4\font\relax}
\providecommand{\BIBforeignlanguage}[2]{{%
\expandafter\ifx\csname l@#1\endcsname\relax
\typeout{** WARNING: IEEEtran.bst: No hyphenation pattern has been}%
\typeout{** loaded for the language `#1'. Using the pattern for}%
\typeout{** the default language instead.}%
\else
\language=\csname l@#1\endcsname
\fi
#2}}
\providecommand{\BIBdecl}{\relax}
\BIBdecl

\bibitem{puy15}
G.~Puy, M.~Davies, and R.~Gribonval, ``Linear embeddings of low-dimensional
  subsets of a hilbert space to $\mathbb{R}^m$,''
  \emph{https://hal.inria.fr/hal-01116153}, 2015.

\bibitem{foucart13}
S.~Foucart and H.~Rauhut, \emph{A Mathematical Introduction to Compressive
  Sensing}, ser. Applied and Numerical Harmonic Analysis.\hskip 1em plus 0.5em
  minus 0.4em\relax Springer New York, 2013.

\bibitem{candes11a}
E.~J. Cand{\`e}s and Y.~Plan, ``Tight oracle inequalities for low-rank matrix
  recovery from a minimal number of random measurements,'' \emph{IEEE Trans.
  Inf. Theory}, vol.~57, no.~4, pp. 2342--2359, 2011.

\bibitem{eftekhari14}
A.~Eftekhari and M.~Wakin, ``New analysis of manifold embeddings and signal
  recovery from compressive measurements,'' \emph{Appl. Comput. Harmon. Anal.},
  2014.

\bibitem{adcock14}
B.~Adcock, A.~C. Hansen, C.~Poon, and B.~Roman, ``Breaking the coherence
  barrier: a new theory for compressed sensing,'' \emph{arXiv:1302.0561}, 2013.

\bibitem{vetterli02}
M.~Vetterli and T.~Blu, ``Sampling signals with finite rate of innovation,''
  \emph{IEEE Trans. Signal Process.}, vol.~50, no.~6, pp. 1417--1428, 2002.

\bibitem{thaper02}
N.~Thaper, S.~Guha, P.~Indyk, and N.~Koudas, ``Dynamic multidimensional
  histograms,'' in \emph{Proc. ACM SIGMOD Int. Conf. Manag. Data}, 2002, pp.
  428--439.

\bibitem{bourrier13}
A.~Bourrier, R.~Gribonval, and P.~P\'{e}rez, ``Compressive gaussian mixture
  estimation,'' in \emph{Proc. IEEE Conf. Acoustics, Speech and Signal
  Processing}, 2013, pp. 6024--6028.

\bibitem{blumensath11}
T.~Blumensath, ``Sampling and reconstructing signals from a union of linear
  subspaces,'' \emph{IEEE Trans. Inform. Theory}, vol.~57, no.~7, pp.
  4660--4671, 2011.

\bibitem{bourrier14}
A.~Bourrier, M.~E. Davies, T.~Peleg, and P.~P\'{e}rez, ``Fundamental
  performance limits for ideal decoders in high-dimensional linear inverse
  problems,'' \emph{IEEE Trans. Inf. Theory}, vol.~60, no.~12, pp. 7928--7946,
  2014.

\bibitem{robinson11}
J.~C. Robinson, \emph{Dimensions, embeddings, and attractors}, ser. Cambridge
  Tracts in Mathematics.\hskip 1em plus 0.5em minus 0.4em\relax Cambridge
  University Press, 2011.

\bibitem{talagrand96}
M.~Talagrand, ``Majorizing measures: the generic chaining,'' \emph{Ann.
  Probab.}, vol.~24, no.~3, pp. 1049--1103, 1996.

\bibitem{clarkson08}
K.~L. Clarkson, ``Tighter bounds for random projections of manifolds,'' in
  \emph{Proc. Symposium on Computational geometry}, 2008, pp. 39--48.

\bibitem{rauhut10}
H.~Rauhut, ``Compressive sensing and structured random matrices,'' in
  \emph{Theoretical Foundations and Numerical Methods for Sparse Recovery},
  ser. Radon Series on Computational and Applied Mathematics, M.~{F}ornasier,
  Ed.\hskip 1em plus 0.5em minus 0.4em\relax De Gruyter, 2010, vol.~9, pp.
  1--92.

\bibitem{dirksen14b}
S.~Dirksen, ``Tail bounds via generic chaining,'' \emph{Electron. J. Probab.},
  vol.~20, no.~53, pp. 1--29, 2015.

\bibitem{cai15}
T.~T. Cai and A.~Zhang, ``Rop: Matrix recovery via rank-one projections,''
  \emph{Ann. Statist.}, vol.~43, no.~1, pp. 102--138, 2015.

\bibitem{chen15}
Y.~Chen, Y.~Chi, and A.~J. Goldsmith, ``Exact and stable covariance estimation
  from quadratic sampling via convex programming,'' \emph{IEEE Trans. Inf.
  Theory}, vol.~61, no.~7, pp. 4034--4059, 2015.

\bibitem{vershynin12}
R.~Vershynin, \emph{Compressed Sensing, Theory and Applications}.\hskip 1em
  plus 0.5em minus 0.4em\relax Cambridge University Press, 2012, ch.
  Introduction to the non-asymptotic analysis of random matrices, pp. 210--268.

\bibitem{baraniuk08}
R.~G. Baraniuk, M.~Davenport, R.~DeVore, and M.~Wakin, ``A simple proof of the
  restricted isometry property for random matrices,'' \emph{Constr. Approx.},
  vol.~28, no.~3, pp. 253--263, 2008.

\bibitem{ayaz14}
U.~Ayaz, S.~Dirksen, and H.~Rauhut, ``Uniform recovery of fusion frame
  structured sparse signals,'' \emph{Appl. Comput. Harmon. Anal.}, vol.~41,
  no.~2, pp. 341--361, 2016.

\bibitem{kueng15}
R.~Kueng, H.~Rauhut, and U.~Terstiege, ``Low rank matrix recovery from rank one
  measurements,'' \emph{Appl. Comput. Harmon. Anal.}, 2015.

\bibitem{bahmani15}
S.~Bahmani and J.~Romberg, ``Sketching for simultaneously sparse and low-rank
  covariance matrices,'' in \emph{Int. Workshop on Computational Advances in
  Multi-Sensor Adaptive Processing}, 2015, pp. 357--360.

\bibitem{achlioptas03}
D.~Achlioptas, ``Database-friendly random projections:
  {J}ohnson-{L}indenstrauss with binary coins,'' \emph{J. Comput. Syst. Sci.},
  vol.~66, no.~4, pp. 671--687, 2003.

\bibitem{bourgain15}
J.~Bourgain, S.~Dirksen, and J.~Nelson, ``Toward a unified theory of sparse
  dimensionality reduction in euclidean space,'' \emph{Geom. Funct. Anal.},
  vol.~25, no.~4, pp. 1009--1088, 2015.

\bibitem{adcock12}
B.~Adcock and A.~C. Hansen, ``A generalised sampling theorem for stable
  rreconstruction in arbitrary bases,'' \emph{J. Fourier Anal. Appl.}, vol.~18,
  no.~4, pp. 685--716, 2012.

\bibitem{dirksen14}
S.~Dirksen, ``Dimensionality reduction with subgaussian matrices: a unified
  theory,'' \emph{Found. Comput. Math.}, pp. 1--30, 2015.

\bibitem{klartag05}
B.~Klartag and S.~Mendelson, ``Empirical processes and random projections,''
  \emph{J. Funct. Anal.}, vol. 225, no.~1, pp. 229--245, 2005.

\bibitem{mendelson07}
S.~Mendelson, A.~Pajor, and N.~Tomczak-Jaegermann, ``Reconstruction and
  subgaussian operators in asymptotic geometric analysis,'' \emph{Geom. Funct.
  Anal.}, vol.~17, no.~4, pp. 1248--1282, 2007.

\bibitem{baraniuk09}
R.~G. Baraniuk and M.~B. Wakin, ``Random projections of smooth manifolds,''
  \emph{Found. Comput. Math.}, vol.~9, no.~1, pp. 51--77, 2009.

\bibitem{mantzel14}
W.~Mantzel and J.~Romberg, ``Compressed subspace matching on the continuum,''
  \emph{Inf. Inference}, 2015.

\bibitem{oymak15}
S.~Oymak, B.~Recht, and M.~Soltanolkotabi, ``Isometric sketching of any set via
  the restricted isometry property,'' \emph{arXiv:1506.03521}, 2015.

\bibitem{riley06}
K.~F. Riley, M.~P. Hobson, and S.~J. Bence, \emph{Mathematical methods for
  physics and engineering}.\hskip 1em plus 0.5em minus 0.4em\relax Cambridge
  University Press, 2006.

\end{thebibliography}

\begin{IEEEbiographynophoto}{\GPlong} is a researcher at Technicolor R\&D France. Before, he was at INRIA in the PANAMA Research team. He obtained the Ph.D. degree in Electrical Engineering from Ecole Polytechnique Fédérale de Lausanne, Switzerland, in January 2014. He also obtained the M.Sc. degree in Electrical and Electronics Engineering from EPFL and the Engineering Diploma (M.Sc.) from Sup\'elec, France, in 2009. He received the EPFL Chorafas Foundation award in 2013 and the EPFL Doctorate award in 2015. 

His research interests are in the field of signal processing, image processing and machine learning. He works on the design of efficient sampling methods for high dimensional data (compressive sampling), low-dimensional representations, and non-linear reconstruction methods using convex and non-convex optimization. His work concerns both theory and applications, mainly in imaging problems. 
\end{IEEEbiographynophoto}

\begin{IEEEbiographynophoto}{\MDlong}(M'00–SM'12–F'15) holds the Jeffrey Collins Chair in Signal and Image Processing at UoE, where he also leads the Edinburgh Compressed Sensing Research Group, and is Head of the Institute for Digital Communications (IDCOM). He received an M.A. in engineering from Cambridge University in 1989 where he was awarded a Foundation Scholarship (1987), and a Ph.D. degree in nonlinear dynamics and signal processing from University College London (UCL) in 1993. He was awarded a Royal Society University Research Fellowship in 1993 and was appointed a Texas Instruments Distinguished Visiting Professor at Rice University in 2012. Currently he also leads the University Defence Research Collaboration (UDRC), a U.K. programme of signal processing research in defence in collaboration with the U.K. Defence Science and Technology Laboratory (DSTL).

His research has focused on nonlinear time series, source separation, sparse representations and compressed sensing. He has made key contributions to these fields including: the development of signal embedding theorems for nonlinear dynamical time series and the development of new theoretical and algorithmic results in Independent Component Analysis (ICA). Most recently he has pioneered the use of sparse representations as a fundamental tool in signal processing, source separation and compressed sensing. This work includes: the proposal and analysis of the highly popular Iterative Hard Thresholding algorithm for sparse reconstruction; the development of a theory for the sampling of structured sparse signal models as well as the development of a variety of new practical sampling and reconstruction techniques that exploit structured sparsity. He has also explored the application of these ideas to advanced medical imaging and RF based sensing applications. He also has an active interest in the related topics of machine learning, high-dimensional statistics and information theory.
\end{IEEEbiographynophoto}

\begin{IEEEbiographynophoto}{\RGlong}(FM'14) is a Senior Researcher with Inria (Rennes, France), and the scientific leader of the PANAMA research group on sparse audio processing. A former student at Ecole Normale Superieure (Paris, France), he received the Ph. D. degree in applied mathematics from Universite de Paris-IX Dauphine (Paris, France) in 1999, and his Habilitation a Diriger des Recherches in applied mathematics from Universite de Rennes I (Rennes, France) in 2007. His research focuses on mathematical signal processing, machine learning, approximation
theory and statistics, with an emphasis on sparse approximation, audio source separation, dictionary learning and compressed sensing. He founded the series of international workshops SPARS on Signal Processing with Adaptive/Sparse Representations. In 2011, he was awarded the Blaise Pascal Award in Applied Mathematics and Scientific Engineering from the SMAI by the French National Academy of Sciences, and a starting investigator grant from the European Research Council.
\end{IEEEbiographynophoto}

\end{document}